\documentclass{aa}

\usepackage[varg]{txfonts}
\usepackage{natbib,graphicx}

\bibpunct{(}{)}{;}{a}{}{,} 

\input{tcilatex}
\begin{document}

\title{Magnetometry: solving the fundamental ambiguity from line pair inversion}

\author{V\'eronique Bommier}

\institute{LESIA, Observatoire de Paris, Universit\'e PSL, CNRS, Sorbonne Universit\'e, Universit\'e de Paris 
\newline 5, Place Jules Janssen, 92190 Meudon, France}

\titlerunning{Magnetometry: solving the fundamental ambiguity}
\authorrunning{V. Bommier}

\date{Received ... / Accepted ...}

\abstract
{All magnetic field vector measurements lead to ambiguous results.}
{We show that observations in two different lines belonging to the same
multiplet but having different absorption coefficients so that they are
formed at two different depths like Fe I 6302.5 \AA\ and 6301.5 \AA ,
enable the resolution of the azimuth ambiguity remaining from the Zeeman
signal interpretation.}
{What is measured by interpretation of the Zeeman effect is the magnetic
field $\vec{H}$, and not the divergence-free magnetic induction $\vec{B}$.
We analyze how the anisotropy of the photosphere,
which is strongly stratified due to gravity and density at the star surface, affects
$\func{div}\vec{H}$ and how the ambiguity resolution has to be performed
in these conditions.}
{As a consequence, two ambiguity-resolved field vector
maps are obtained at two different but close altitudes, which enable the
derivation of the current density full vector via $\func{curl}\vec{H}=
\vec{J}$. This reveals the horizontal component of the current density,
which is generally found markedly larger than the better known vertical one. We
observe some systematical trends, of which we present examples in the paper, 
like circular currents wrapping spots
clockwise about a positive polarity spot and anticlockwise about a negative
polarity spot and strong horizontal current components crossing active
region neutral lines.}
{We finally remark that the Na I D1 and D2 lines form another such line
pair. We propose them as an access to the low chromosphere
where they are formed. However, a better spatial resolution for both 
observations and analysis 
would probably be necessary in such a medium.}

\keywords{Magnetic fields -- Polarization -- Sun: magnetic fields -- Sun: photosphere -- Sun: chromosphere -- Sun: magnetic topology}

\offprints{V. Bommier, \email{V.Bommier@obspm.fr}}

\maketitle

\section{Introduction}

In solar atmosphere, magnetic field and electric currents are
strongly coupled via the Amp\`{e}re's circuital law. Moreover, and without
entering in details, the presence of strong electric currents, for instance
under the form of a current sheet, could be the place where an eruptive
event could occur. Thus, the knowledge of the electric current is essential
for understanding the atmosphere stability and evolution for space
weather purposes. However, the electric current vector components result
from spatial derivatives of the magnetic field vector components via
the Maxwell equation $\func{curl}\vec{H}=\vec{J}$. This makes them
particularly difficult to derive from the observations because this requires
a good accuracy for the primitive quantity (in the mathematical meaning), 
which is the magnetic field. This requires also that the magnetic
field vector is fully known without any ambiguity. The
Zeeman effect observation is ambiguous in terms of magnetic field vector 
because two field vectors symmetrical with respect to the
line-of-sight, i.e., which have the same longitudinal component but
opposite transverse components, are responsible for the same polarization.
They are not discriminated at the spectropolarimetric data inversion step.
This ambiguity is often referred to as the azimuth ambiguity, or $180%
{{}^\circ}%
$ ambiguity, or fundamental ambiguity.

The derivation of the electric current full vector requires that the
ambiguity is solved after the spectropolarimetric data inversion. A
review of different methods can be found in \citet{Metcalf-etal-06}.
However, sunspots offer the opportunity of an intuitive solution of the
ambiguity in their penumbra from the spot magnetic polarity. For a
negative spot, the penumbra field vectors point towards the spot center,
whereas for a positive spot the penumbra field vectors diverge from the spot
center.\ Some authors took advantage of this opportunity to obtain current
density vectors in spot penumbrae. \citet{SocasN-05} applied the SIR code %
\citep{RuizC-delToro-92} for the spectropolarimetric data inversion. This
code uses several depth nodes, where the magnetic field is independently
determined, which enables the derivation of the field variation along depth. 
The ambiguity intuitive solution was further introduced in the spot
penumbra. Then, the full current density vector was derived in the
spot. The same method was applied to HINODE/SOT/SP data by %
\citet{Puschmann-etal-10b}. The quality of HINODE data permitted to
show a coherence between the filamentary structure of the penumbra %
and the current density vector lines. Significant currents are seen to wrap
around the hotter, more elevated regions with lower and more horizontal
magnetic fields that harbor strong upflows and radial outflows (the
intraspines).

Indeed, the derivation of the three coordinates of the current density
vector requires the knowledge of the depth variations of the magnetic field,
which is rarely achieved. In general, a single line is observed, which is
formed at a given depth. Except when the SIR code or a code of
similar technique is applied, the magnetic field is derived at this single
depth and only the vertical component $J_{z}$ of the current density can be
finally derived.\ Thus, full vector current density determinations remain
rare. We present below such determinations, which are not based on the SIR
code application, but on two line observations and single-line
Milne-Eddington inversions. The inversion method is described in section \ref%
{subsect--inversion}. The depth difference in line formation altitude was
previously directly derived from other HINODE/SOT/SP\ data %
\citep{Faurobert-etal-09}. This enables the complete derivation of $\func{div%
}\vec{H}$ from the observations, which enables the ambiguity resolution, 
the method of which is described in section \ref{subsect--ambiguity}%
. In particular, we analyze how the anisotropy of the photosphere,
which is strongly stratified due to gravity and density at the star surface,
affects $\func{div}\vec{H}$ and how the ambiguity resolution has
to be performed in these conditions. We present example results in section %
\ref{sect--vector-maps}. As \citet{Puschmann-etal-10b}, we obtain that the
horizontal component of the current density vector is much stronger than the
vertical one. The vertical component is the most frequently published and
discussed. The horizontal component remained generally unknown, which is yet
much larger. We discuss examples of typical structures we derived for the
horizontal component, in sunspot penumbrae and across some neutral lines. 
Obviously, applying the SIR code inversion \citep{RuizC-delToro-92} 
would provide more refined vertical variations for evaluating divergences 
and electric currents.

We tested our multiline ambiguity solution method on different spectral
lines observed with the TH\'{E}MIS\ telescope operating in its new multislit
mode.\ The two splitted beams enter the spectrograph by two parallel slits. 
As a result, we obtained that the pair of lines must belong to the same multiplet
in order to be formed in the same manner, but with different absorption
coefficients. This is the case of the Fe I 6301.5 \AA\ and 6302.5\ \AA\ line
pair. With these conditions, their depth formation difference remains rather
constant as visible in the simulation by %
\citet{Khomenko-Collados-07}. This line pair is interestingly the one
observed by HINODE/SOT/SP. In sections \ref%
{subsect--THEMIS1} and \ref{subsect--THEMIS2}, we present 
the magnetic field maps observed
in a sunspot penumbra and above a neutral line. 

It has to be recalled that what is measured by interpretation of the
Zeeman effect is the magnetic field $\vec{H}$, and not the magnetic
induction $\vec{B}$  \citep{Bommier-20}. The magnetic induction $\vec{B}$ is
divergence-free, when the magnetic field $\vec{H}$
is not. They are related by $\vec{B}=\mu_0(\vec{H}+\vec{M})$, where 
$\vec{M}$ is the magnetization. However, in usual solar models, the
reported electron density is such that the modulus of $\vec{M}$ is very small, 
with respect to the modulus of $\vec{H}$,
which results in $\vec{B}\approx\mu_0\vec{H}$, which makes 
$\vec{H}$ quite divergence-free also. In this respect, minimizing 
$\func{div}\vec{H}$ makes sense for solving the ambiguity. However,
the Sun's surface electron density could be much higher due to
electron thermal escape in the solar interior, where the electron
thermal velocity overpasses the star gravitation escape velocity
by a factor of 14 and the proton escape velocity by a factor of 6,
as remarked by \citet{Bommier-20}. The model electron density
is in fact derived from ionization equilibria modelled from the spectrum,
and in addition within the local electric neutrality hypothesis 
\citep[Section V]{Vernazza-etal-73}, which 
has to be questioned in the presence of the star gravitation field
as explained above. Thus,  $\func{div}\vec{H}$ could be non-zero
as observed \citep[see the review by][]{Balthasar-18}, 
but its minimization could be kept as a method
for solving the fundamental ambiguity. The results presented in the 
present paper confirm this approach. 

In practice, we will apply Gauss 
units for describing the magnetic field. They are in fact units for
$\mu_0 \vec{H}$.

\section{Data Analysis First Step: Milne-Eddington Inversion}

\label{subsect--inversion}

We applied the Milne-Eddington inversion code UNNOFIT. Based on the
Unno-Rachkovsky solution of the transfer equation for the radiation Stokes
parameters of a spectral line, the code accounts for all the
magneto-optical effects \citep{Landolfi-Landi-82}.\ The code applies the
Levenberg-Marquardt algorithm to the theoretical profile to make it fitting
the observed profile.\ Eight parameters were entered in the algorithm, which
are namely: the line center frequency $\lambda _{0}$, the Doppler width $%
\Delta \lambda _{\mathrm{D}}$, $\eta _{0}$, the three magnetic field
coordinates, the Voigt parameter $a$, and the ratio $B_{1}/B_{0}$ of the two
parameters characterizing the Milne-Eddington atmosphere.\ This was the
original code developed by \citet{Landolfi-etal-84}.

However, since the pioneering work by \citet{Stenflo-73}, it is
well known that the atmosphere is permeated by unresolved magnetic
structures, which can be roughly modelled with a magnetic filling factor $%
\alpha $. Following a wish by Egidio Landi Degl'Innocenti, we introduced
such a magnetic filling factor in his UNNOFIT code.
We tested the code and validated it \citep{Bommier-etal-07}, but we obtained
that the unresolved magnetic field strength $H$ and the magnetic filling
factor $\alpha $ cannot be separately determined (see Fig. 4 of that paper). 
$H$\ is here the modulus of the magnetic field vector, which is
different from its longitudinal component. Only their product $\alpha H$,
which is the local average magnetic field strength, is obtained as final
result. This occurs when the field is not strong enough to well separate the
Zeeman components with respect to the Doppler width (typically 1000 G at
least).\ This is the effect of the larger number of parameters to be
determined with respect to the number of independent parameters
provided by the line observation. A line is characterized by a position, a
width, a central depth, and three Stokes parameters, which enable
the calculation of the three magnetic field components. This
results in only 6 independent parameters. The Voigt $a$ parameters
characterizes the far wings and plays a specific role there.\ Thus, with the
filling factor, the number of searched for parameters exceeds the number of
available parameters.\ However, as shown by the tests, the local
average magnetic field stregth $\alpha H$\ is finally well determined. Later
on, a tenth parameter was introduced in the inversion, which is the velocity
gradient along the line formation depth, which depends on an additional
observed parameter, which is the line asymmetry \citep{Molodij-etal-11}.

The inversion is performed in the four Stokes profiles simultaneously. 
The same weight of
unity was given to the four Stokes parameters for the chi-square calculation 
because $I$\ plays an important role for the determination of the magnetic filling factor %
$\alpha $\ (usually, $I$ is less weighted). It is
well-known that in weak fields the linear polarization
Stokes parameters $(Q,U)$ depend quadratically on the transverse magnetic
field, whereas the circular polarization Stokes parameter $V$\ depends
linearly on the longitudinal magnetic field. As the chi-square of
the four Stokes parameter are simply added in the algorithm, the
longitudinal and transverse fields are simultaneously determined in a unique
procedure.  This results in quite comparable accuracies. We plotted the
difference histograms for the cartesian components 
$\mu_0 H_{x},\mu_0 H_{y},\mu_0 H_{z}$ in
the line-of-sight reference frame assumed to be at disk-center for
the same test data as above. We obtained quite comparable
widths or inaccuracy of 15 G for $\mu_0 H_{x}$ and $\mu_0 H_{y}$, and 10 G for $%
\mu_0 H_{z}$, for $\alpha\mu_0 H>10$ G, from noised profiles 
at the $1.5\times 10^{-3}$ level for the polarimetric noise, which is
not far from the HINODE/SOT/SP one that we evaluated at $1.2\times 10^{-3}$
in profile far wings ($1.0\times 10^{-3}$\ during our last TH\'{E}MIS
campaigns described in this paper).

Although the inversion finally provides only the $\alpha\mu_0 H$ product, it is 
important to take $\alpha $ into account during the
inversion for a good determination of the field inclination \citep{Leka-etal-22}. Due to the
sensitivity difference between the longitudinal and transverse fields,
forcing $\alpha =1$\ (i.e., ignoring $\alpha $) would lead to bad
inclinations in regions where $\alpha $\ is far from unity.

\citet{Landolfi-etal-84} developed the light UNNOFIT version for the normal
Zeeman triplet, and the heavier UNNOFIT2 version specific for Zeeman
multiplets. More details are given in \citet{Bommier-13}.\ We were thus able
to treat any kind of line.

\section{Warning about the aspect ratio applied for the ambiguity resolution}

\label{subsect--warning}

In the present paper, we present an ambiguity resolution method, 
in which we apply a scaling based on the aspect ratio of the stratified atmosphere
(see below Subsect. \ref{subsect--aspect}). Our intuition is based on the non-zero
value generally observed for the magnetic field divergence, with vertical
magnetic field gradients on the order of 3 G/km, when the horizontal gradient
is on the order of 0.3 G/km only \citep[see the review by][]{Balthasar-18}.
However, it is important to advertise that the determination of these gradients
depends on two things: first, the in-depth analysis, which requires appropriate 
inversion codes like the SIR code \citep{RuizC-delToro-92}, for 
vertical gradient, and second, a refined spatial resolution, for 
horizontal gradient. And, third, these gradients may also depend
on the studied solar structure.

In the present paper,
we did not apply the SIR or similar inversion code. We applied rougher
methods. And the observation spatial resolution is not the better one.
For instance, \citet{Puschmann-etal-10a}
studied in details a sunspot penumbra with a better spatial resolution,
and they obtain smaller gradient values, on the order of 0.25 G/km
for the vertical gradient (see their Fig. 8). However, their horizontal
gradient is about 0.05 G/km only, which remains significantly different 
from the vertical gradient. They applied the SIR inversion code.
\citet{Buehler-etal-15} use observation
with similar spatial resolution as ours, and find then a vertical gradient
similar to ours. They apply the SPINOR inversion code \citep{Frutiger-etal-00}.
They do not strictly find a magnetic flux conservation,
but they can finally impose the magnetic flux conservation to their results.
When done, the gradient of their inversion agrees with the thin-tube approximation
(see the end of their Sect 4.2). They interestingly distinguish between
core and canopy pixels (see their Fig. 3), which have different signs
for their vertical magnetic field gradients. This underlines the effect
of the spatial resolution.

Below is a non-exhaustive list of some effects that have been neglected 
when considering the vertical magnetic field gradient in sunspot umbrae, in our work:

\begin{itemize}
\item the presence of velocity-temperature or magnetic field-temperature 
(or continuum intensity) correlations, which could be produced by, e.g., umbral dots 
at below the spatial resolution of the observations;
\item blending of the atomic 
spectral lines used to measure H in umbrae by molecular lines (often unidentified, 
or with poor molecular data);
\item the effect of straylight from the penumbrae, from surrounding plage 
and from the quiet Sun. Straylight can play a major role in the umbra by affecting 
spectral lines significantly (it is not sufficient to model it via a simplistic filling factor, 
as we do);
\item departure from a purely hydrostatic equilibrium (e.g., by including the 
$\vec{J} \times \vec{B}$ force in the force-balance), which can affect the line formation 
height;
\item the effect of a non-horizontal and corrugated $\tau = 1$ surface, 
which is strongly inclined in most parts of the Sun, including in the umbra. 
This can falsify the horizontal gradients of the field deduced from observations 
with finite spatial resolution;
\item non-LTE effects in spectral lines, which can also change their 
formation heights.
\end{itemize}

Consequently, as long as these and other effects have not been properly tested 
in realistic model atmospheres and with state-of-the-art radiative transfer computations, 
the scaling factor that we introduce below must be considered to be a purely empirical factor. 
This factor will depend on the quality of the data and of the inversion as well as 
on the solar feature being studied.

\section{Data Analysis Second Step: Azimuth Ambiguity Resolution}

\label{subsect--ambiguity}

We applied a modification to the ME0\ code developed by Metcalf, Crouch,
Barnes, \& Leka and now available on the web\footnote{%
http://www.cora.nwra.com/AMBIG/} \citep{Leka-etal-09}. This code applies the
\textquotedblleft Minimum Energy\textquotedblright\ Method initially
described by \citet{Metcalf-94}, which consists in searching for
the field vector orientation that minimizes $\left\vert \func{div}\vec{H}%
\right\vert +\lambda \left\vert J_{Z}\right\vert $, where $\vec{J}$
is the current density vector, $J_{Z}$ its vertical component, and $\lambda $
a positive weight parameter usually fixed at unity 
(with $\mu_0\vec{H}$ expressed in Gauss). Minimizing $\left\vert 
\func{div}\vec{B}\right\vert $ is a natural requirement imposed by Maxwell's
equations. As explained in the Introduction, we will extent this requirement
to what is really measured, i.e., $\left\vert\func{div}\vec{H}\right\vert $.
On the other hand, minimizing the current density minimizes the
maximum allowed free magnetic energy \citep{Metcalf-94}. The minimization is
performed globally on the whole map by applying the \textquotedblleft
simulated annealing\textquotedblright\ algorithm. Its application
to this minimization problem is described by \citet{Crouch-etal-09}. The
method is then complemented by propagating the solution via the acute angle
method below a certain field strength threshold \citep{Leka-etal-09} %
presently taken at 400 G.\ The acute angle method consists in selecting
from two ambiguous solutions symmetrical with respect to the line-of-sight,
the solution that makes an acute angle (in the transverse plane)
with the vector to be compared. In the original ME0 method, one single map
is used and the vertical derivatives of the magnetic field are derived from
a current-free reconstruction of the magnetic field. Reconstructions based
on less restrictive hypotheses were later on introduced by this method
authors \citep{Metcalf-etal-06,Leka-etal-09}. As the field vector ambiguity
concerns its transverse component in the line-of-sight coordinates, whereas
the reconstruction is performed in the heliographic coordinates, back and
forth transformations have to be performed between these two systems of
coordinates.

More precisely, $\func{div}\vec{H}$ is calculated in the heliographic
reference frame in the original ME0 code because $\partial
H_{z}/\partial z$ is evaluated in this frame by the reconstruction.\ Our
approach avoids the vertical reconstruction of the field by the introduction
of two maps recorded in two lines formed at two different depths. In our
method we kept the minimization procedure of ME0 but we calculated $\func{div%
}\vec{H}$\ in the line-of-sight coordinates instead, which was
possible from our 2-line observations. We accordingly modified the
corresponding subroutine of ME0. We calculated the expression of $\func{div}%
\vec{H}$\ adapted to the case of a line formation plane inclined with
respect to the line-of-sight in order to develop a method able to
treat maps of any location on the solar disk. Below and in Appendix \ref%
{Appendix}, we describe this calculation. A different calculation, %
which also accounts for the case of a line-of-sight inclined with respect
to the local vertical, was developed by %
\citet{Crouch-etal-09,Crouch-13,Crouch-15}. We have first to introduce the
various reference frames entering the calculation.

\subsection{Reference frames}

Our approach is in agreement with \citet{AllenG-Hagyard-90} in the
limit $P=0$. In the following, we denote as $Oxyz$\ the
line-of-sight (l.o.s.) reference frame and $OXYZ$\ the heliographic one at
the observed region location. The l.o.s. reference frame is defined by $Oz$\
being the l.o.s. itself oriented towards the observer, and $Oy$\ being
parallel to the disk central meridian oriented towards the solar north. The
heliographic reference frame has $OX$\ aligned with the local parallel solar
west oriented, $OY$\ aligned with the local meridian solar north oriented,
and $OZ$\ along the solar radius oriented from the Sun's center towards
outside. We denote as $\vec{R}$\ the unit vector along the $OZ$\ axis of the
heliographic reference frame, which is also the local solar radius.
We denote with indexes $(x,y,z)$ the vector coordinates in the l.o.s.
reference frame, and $(X,Y,Z)$ those in the heliographic reference frame. %
\citet{AllenG-Hagyard-90} denote with the upper index $l$ the vector
coordinates in the l.o.s. reference frame, and $h$ those in the heliographic
reference frame. In the l.o.s. reference frame, the $\vec{R}$\
coordinates are%
\begin{equation}
\left\{ 
\begin{array}{l}
R_{x}=\sin (L-L_{c})\cos b \\ 
R_{y}=\sin b\cos b_{0}-\cos (L-L_{c})\cos b\sin b_{0} \\ 
R_{z}=\sin b\sin b_{0}+\cos (L-L_{c})\cos b\cos b_{0}%
\end{array}%
\right. \ ,  \label{R-vector}
\end{equation}%
where $L$\ and $b$\ are, respectively, the longitude and latitude
of the center $O$\ of the observed region, $L_{c}$\ is the disk central
meridian longitude, and $b_{0}$\ is the disk center latitude. The
other unit vectors of the $OXYZ$\ heliographic reference frame are denoted
as $\vec{I}$\ along $OX$\ and $\vec{K}$\ along $OY$, with 
\begin{equation}
\left\{ 
\begin{array}{l}
I_{x}=\cos (L-L_{c}) \\ 
I_{y}=\sin (L-L_{c})\sin b_{0} \\ 
I_{z}=-\sin (L-L_{c})\cos b_{0}%
\end{array}%
\right.
\end{equation}%
and%
\begin{equation}
\left\{ 
\begin{array}{l}
K_{x}=-\sin (L-L_{c})\sin b \\ 
K_{y}=\cos b\cos b_{0}+\cos (L-L_{c})\sin b\sin b_{0} \\ 
K_{z}=\cos b\sin b_{0}-\cos (L-L_{c})\sin b\cos b_{0}%
\end{array}%
\right.
\end{equation}%
in the l.o.s. reference frame. These formulae are obtained by applying 3
rotations successively to transform the heliographic reference frame into
the l.o.s. one, namely: 1/ rotation of $b$\ about the $OX$\ %
axis; 2/ rotation of $-(L-L_{c})$\ about the new $OY$\ axis; 3/
rotation of $-b_{0}$\ about the new $Ox$\ axis. Conversely, the coordinates
of the l.o.s. basic unit vectors $\vec{i}$, $\vec{k}$\ and $\vec{\ell}$\
(l.o.s. vector) in the heliographic reference frame are%
\begin{equation}
\left\{ 
\begin{array}{l}
i_{X}=\cos (L-L_{c}) \\ 
i_{Y}=-\sin (L-L_{c})\sin b \\ 
i_{Z}=\sin (L-L_{c})\cos b%
\end{array}%
\right.  \label{equation-i}
\end{equation}%
and%
\begin{equation}
\left\{ 
\begin{array}{l}
k_{X}=\sin (L-L_{c})\sin b_{0} \\ 
k_{Y}=\cos b\cos b_{0}+\cos (L-L_{c})\sin b\sin b_{0} \\ 
k_{Z}=\sin b\cos b_{0}-\cos (L-L_{c})\cos b\sin b_{0}%
\end{array}%
\right.  \label{equation-k}
\end{equation}%
and%
\begin{equation}
\left\{ 
\begin{array}{l}
\ell _{X}=-\sin (L-L_{c})\cos b_{0} \\ 
\ell _{Y}=\cos b\sin b_{0}-\cos (L-L_{c})\sin b\cos b_{0} \\ 
\ell _{Z}=\sin b\sin b_{0}+\cos (L-L_{c})\cos b\cos b_{0}%
\end{array}%
\right. \ .  \label{equation-l}
\end{equation}%
The transformation from the coordinates $\left( x,y,z\right) $\ 
of a given vector in the l.o.s. reference frame into its
coordinates $\left( X,Y,Z\right) $\ in the heliographic reference frame, can
be written as%
\begin{equation}
\left\{ 
\begin{array}{l}
X=xi_{X}+yk_{X}+z\ell _{X} \\ 
Y=xi_{Y}+yk_{Y}+z\ell _{Y} \\ 
Z=xi_{Z}+yk_{Z}+z\ell _{Z}%
\end{array}%
\right.  \label{lostosolar}
\end{equation}%
and conversely%
\begin{equation}
\left\{ 
\begin{array}{l}
x=XI_{x}+YK_{x}+ZR_{x} \\ 
y=XI_{y}+YK_{y}+ZR_{y} \\ 
z=XI_{z}+YK_{z}+ZR_{z}%
\end{array}%
\right. \ .  \label{solartolos}
\end{equation}

\subsection{Deprojection}

A map is obtained from a scan of the spectrograph slit along the
solar image. The map is reconstructed by positioning side by side all the
slit outputs. The mapped quantities are then given on a rectangular system
of pixels in the l.o.s. reference frame, the $\Delta x$\ pixel size along $%
Ox $\ being given by the scan step size and the $\Delta y$\ pixel size along 
$Oy $\ being given by the camera pixel size. Thus $\Delta x\neq \Delta y$\
in general, which results in an anamorphosis of the map. Once the ambiguity
is solved and a single magnetic field vector is obtained for each pixel, its
coordinates may be transformed into the heliographic reference frame. The
deprojection of the pixel array is a more complicated task because the meshs
are no more rectangular in general, as described for instance in %
\citet{AllenG-Hagyard-90}. In the present paper and related studies, %
we have roughly approximated a rectangular shape for the deprojected map,
however, with pixel side sizes of $\Delta X=\Delta x/I_{x}$\ and $\Delta
Y=\Delta y/K_{y}$. This is exact when the map center is located on the solar
equator and when $b_{0}=0$. This departs from exactitude when the latitudes
of the disk and/or map centers depart from 0. The cosine of the heliocentric
angle $\theta $\ is $\mu =R_{z}$.

\subsection{The quantity to be minimized}

\label{subsect-divrot}

\begin{figure}
\includegraphics[width=8.8cm]{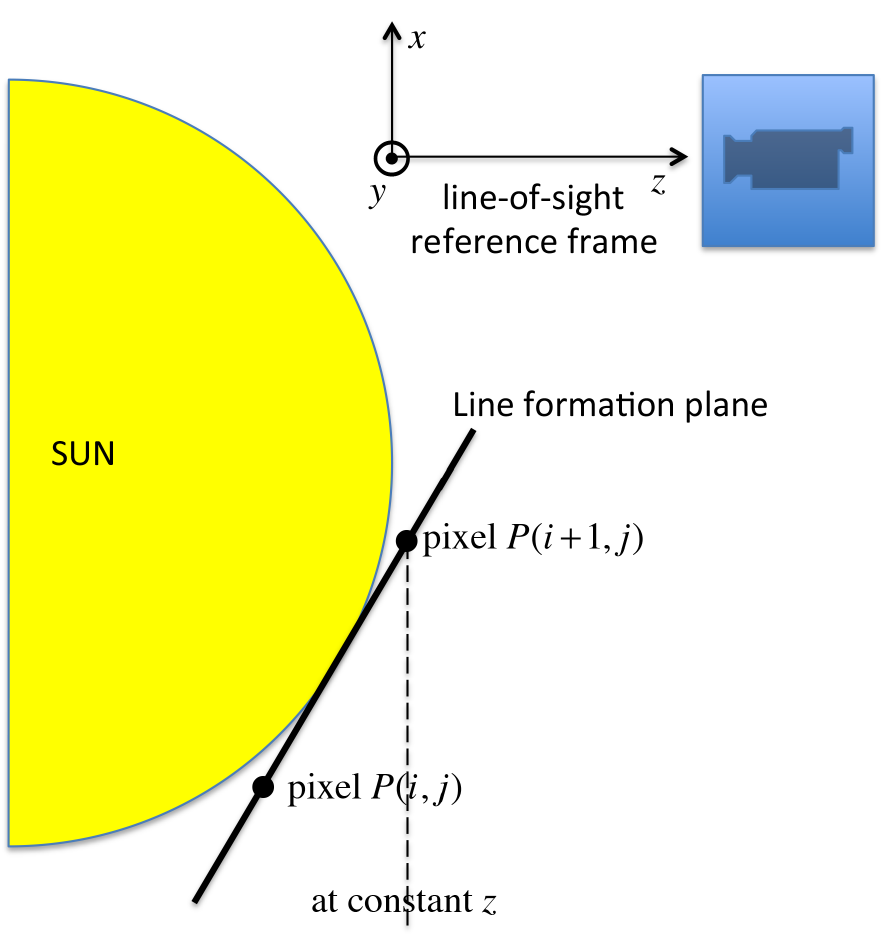}
\caption{$\Delta_{x}H_{x}^{(m)}$ is the difference between the $H_{x}$ components measured in pixels $P(i+1,j)$ and $P(i,j)$. It can be seen that the $z$ coordinate is not the same for the two pixels for an observation out of disk center, which is in contradiction with the definition of the partial derivative along $x$.}
\label{partiald}
\end{figure}

As stated above, we applied the \textquotedblleft simulated
annealing\textquotedblright\ algorithm to globally minimize $\left\vert 
\func{div}\vec{H}\right\vert +\lambda \left\vert J_{Z}\right\vert $.
The parameter $\lambda $\ was taken as unity
(with $\mu_0\vec{H}$ expressed in Gauss). Indeed, in his pioneering
paper, \citet{Metcalf-94} recommends to minimize the alternative quantity $%
\left\vert \func{div}\vec{H}\right\vert +\lambda \left\Vert \vec{J}%
\right\Vert $\ as the physically meaningful quantity for \textquotedblleft
energy minimization\textquotedblright , with the full current vector and not
only its $Z$-component.\ We first tried this minimization because we had
also the transverse current component (within the ambiguity) from our 2-line
observations, but it did not converge correctly. On the contrary,
we found that the algorithm sucessfully worked out, when %
we\ instead minimized $\left\vert \func{div}\vec{H}\right\vert +\lambda \left\vert J_{Z}\right\vert $\ with $\lambda =1$. We tried
various $\lambda $\ values and also found $\lambda =1$\ to be the best
choice, which leads to maps at most free of articifial
defaults like strong current lines delineating zones of different ambiguous
solution choice by the algorithm, or with minimal checkered zones where the
ambiguity resolution failed.\ The minimization of $\left\vert \func{%
div}\vec{H}\right\vert +\lambda \left\vert J_{Z}\right\vert $ was
later performed also by Metcalf's team \citep{Metcalf-etal-06}.

However, the map under study is usually not located at
disk center, which results in a certain inclination of the map formation
plane with respect to the l.o.s. referred to by the heliocentric
angle $\theta $\ and its cosine $\mu =\cos \theta $. We denote as
\textquotedblleft map formation plane\textquotedblright\ the heliographic
horizontal plane, i.e., the plane perpendicular to the local solar
radius. The map results from interpretation of polarization in spectral
lines. As explained below in Sect. \ref{subsect-depthdiff}, the radiation
received along the line-of-sight in a given spectral line comes essentially
from a certain height in the solar atmosphere (this is the Eddington-Barbier
approximation), which defines an heliographic horizontal plane, which
contains the location of the points where the magnetic field is measured by
spectropolarimetry. This plane is not perpendicular to the
line-of-sight, when the observed region is not located at disk center.

But the divergence has to be computed in a rectangular reference frame.\ For
doing this, there are two possibilities. Either the coordinates of the two
ambiguous magnetic field vectors obtained in the line-of-sight reference
frame and symmetrical with respect to the line-of-sight, which are
of the $H^{(l)}$ type, are transformed into heliographic
coordinates of the $H^{(h)}$ type of \citet{AllenG-Hagyard-90}. The
ambiguity is then resolved in the heliographic reference frame,
where the measurements are all located in the horizontal plane perpendicular
to the $OZ$ axis. This is the method applied by \citet{Leka-etal-09} in
their ME0 code described above. Or, as for us, we resolved
the ambiguity in the line-of-sight coordinates of the field as
obtained from the measurements, which are of the $H^{(l)}$ type of %
\citet{AllenG-Hagyard-90}. In this case, the measurements
are located along a plane inclined with respect to the
line-of-sight, which is the $Oz$\ axis. This inclination 
described below and in Fig. \ref{partiald} is at the origin of
balancing terms in the divergence and curl expressions in terms of
line-of-sight or $H^{(l)}$-type magnetic field vector coordinates, as
derived in Appendix \ref{Appendix}.

In Appendix \ref{Appendix} we derive the expressions of $\func{div}\vec{H}$, 
$J_{Z}$\ and $J_{x,y}$ as a function of the l.o.s. reference frame magnetic
field coordinates, when the fields are measured along an inclined formation
plane perpendicular to the local solar radius unit vector $\vec{R}$\ of
l.o.s. components $R_{x,y,z\text{ }}$given in Eq. (\ref{R-vector}). We 
obtain%
\begin{eqnarray}
\func{div}\vec{H} &=&\frac{\Delta _{x}H_{x}^{(m)}}{\Delta x}+\frac{\Delta
_{y}H_{y}^{(m)}}{\Delta y}+\frac{\Delta _{z}H_{z}^{(m)}}{\Delta z}
\label{divergence} \\
&&+\frac{R_{x}}{R_{z}}\frac{\Delta _{z}H_{x}^{(m)}}{\Delta z}+\frac{R_{y}}{%
R_{z}}\frac{\Delta _{z}H_{y}^{(m)}}{\Delta z}\ .  \nonumber
\end{eqnarray}%
We denote as $\vec{H}^{(m)}$\ the vector measured along the inclined %
plane, where the line is formed. $Oxyz$\ is the l.o.s. reference frame 
and $Oy$\ is solar north oriented. $\Delta
_{x}H_{x}^{(m)}$\ and $\Delta _{y}H_{y}^{(m)}$\ are respectively
the difference of the measured component $H_{x}^{(m)}$or $H_{y}^{(m)}$\
between two neighboring pixels in $x$\ or $y$\ direction. $\Delta x$\ or $%
\Delta y$\ denotes the distance between the neighboring pixels in
the \textquotedblleft sky plane\textquotedblright , which is the
plane perpendicular to the l.o.s. at the location of the observed region.
The magnetic field may eventually be averaged between the two lines. If this
average has to be performed before the ambiguity is solved, the acute angle
method is applied between the two lines to select the ambiguous solutions to
average. This is a reasonable approximation, which saves computation time in
the simulated annealing. This corresponds to assume that the field lines do
not twist so much. As for $\Delta _{z}H_{z}^{(m)}$, it is the difference
between the values obtained from the two lines with a difference in
depth formation $\Delta z$\ along the l.o.s.. Explicit definitions
of all these quantities are given in Appendix \ref{Appendix}. The way of
calculating these quantities is detailed at the beginning of the Appendix.
When the map center is located on the solar equator and when $b_{0}=0$, the
above Eq. (\ref{divergence}) simplifies in%
\begin{eqnarray}
\func{div}\vec{H} &=&\frac{\Delta _{x}H_{x}^{(m)}}{\Delta x}+\frac{\Delta
_{y}H_{y}^{(m)}}{\Delta y}+\frac{\Delta _{z}H_{z}^{(m)}}{\Delta z}
\label{divergence-equator} \\
&&+\tan \theta \frac{\Delta _{z}H_{x}^{(m)}}{\Delta z}  \nonumber
\end{eqnarray}%
where $\theta $\ is the heliocentric angle (positive for the West
side of the central meridian and negative for the East side). The
second line of these equations accounts for the $z$\ variation hidden in $%
\Delta _{x}H_{x}^{(m)}/\Delta x$\ and $\Delta _{y}H_{y}^{(m)}/\Delta y$,
because $H^{(m)}$\ is measured along the line formation plane which is not
perpendicular to $Oz$\ in the general case.

In other words, if one considers two neighboring pixels, referred
to as $P\left( i,j\right) $\ and $P\left( i+1,j\right) $, separated by the
length $\Delta x$\ in the plane perpendicular to the line-of-sight, and if
one considers a quantity $A$\ measured via the same line at the two places, 
namely $A_{i,j}$\ and $A_{i+1,j}$, the ratio $(A_{i+1,j}-A_{i,j})/%
\Delta x=\Delta _{x}A/\Delta x$ is not an approximate value of the
partial derivative $\partial A/\partial x$ because $\Delta _{x}A$ 
also\ involves a variation along the line-of-sight $z$ %
together with the variation in $x$, when the observation is not performed
at disk center,\ i.e., when the line formation plane is not
perpendicular to the line-of-sight. From the definition, $\partial
A/\partial x$\ has to be evaluated at constant $y$\ and $z$. In the
evaluation of $\Delta _{x}A=A_{i+1,j}-A_{i,j}$, one has $z\left[ P(i+1,j)%
\right] \neq z\left[ P(i,j)\right] ,$ as it is visible in Fig. \ref%
{partiald}, when the line formation plane is inclined with respect to the
line-of-sight, which is also the $Oz$ axis. As a consequence, $z$\
is not kept constant in the partial derivative numerical evaluation, when
the observation is done out of disk center. This was the reason to perform
the Appendix \ref{Appendix} calculations. 

For numerical computation, the divergence was computed at the center of a
pixel, by averaging between the variations along each side of the pixel. 
We use two lines formed at two different depths. This
results in%
\begin{eqnarray}
\Delta _{x}H_{x}^{(m)} &=&(H_{x}^{(1)}(i+1,j)-H_{x}^{(1)}(i,j) \\
&&+H_{x}^{(1)}(i+1,j+1)-H_{x}^{(1)}(i,j+1)  \nonumber \\
&&+H_{x}^{(2)}(i+1,j)-H_{x}^{(2)}(i,j)  \nonumber \\
&&+H_{x}^{(2)}(i+1,j+1)-H_{x}^{(2)}(i,j+1))/4\ ,  \nonumber
\end{eqnarray}%
\begin{eqnarray}
\Delta _{y}H_{y}^{(m)} &=&(H_{y}^{(1)}(i,j+1)-H_{y}^{(1)}(i,j) \\
&&+H_{y}^{(1)}(i+1,j+1)-H_{y}^{(1)}(i+1,j)  \nonumber \\
&&+H_{y}^{(2)}(i,j+1)-H_{y}^{(2)}(i,j)  \nonumber \\
&&+H_{y}^{(2)}(i+1,j+1)-H_{y}^{(2)}(i+1,j))/4\ ,  \nonumber
\end{eqnarray}%
\begin{eqnarray}
\Delta _{z}H_{z}^{(m)} &=&(H_{z}^{(2)}(i,j)-H_{z}^{(1)}(i,j) \\
&&+H_{z}^{(2)}(i+1,j)-H_{z}^{(1)}(i+1,j)  \nonumber \\
&&+H_{z}^{(2)}(i,j+1)-H_{z}^{(1)}(i,j+1)  \nonumber \\
&&+H_{z}^{(2)}(i+1,j+1)-H_{z}^{(1)}(i+1,j+1))/4\ ,  \nonumber
\end{eqnarray}%
where the indexes $(1)$ and $(2)$\ correspond to
\textquotedblleft line 1\textquotedblright\ and \textquotedblleft line
2\textquotedblright .

For the current density vector component $J_{Z}$\ along the direction
perpendicular to the inclined plane we obtain (curl of the magnetic
field)%
\begin{eqnarray}
J_{Z} &=&R_{z}\left[ \frac{\Delta _{x}H_{y}^{(m)}}{\Delta x}-\frac{%
\Delta _{y}H_{x}^{(m)}}{\Delta y}\right]  \label{rotationnel} \\
&&+R_{x}\frac{\Delta _{y}H_{z}^{(m)}}{\Delta y}-R_{y}\frac{\Delta
_{x}H_{z}^{(m)}}{\Delta x}\ .  \nonumber
\end{eqnarray}%
It is possible to obtain analogously the two components of the current
density vector in the plane perpendicular to the line-of-sight (see Appendix %
\ref{Appendix})%
\begin{eqnarray}
J_{x} &=&\left[ \frac{\Delta _{y}H_{z}^{(m)}}{\Delta y}-\frac{\Delta
_{z}H_{y}^{(m)}}{\Delta z}\right] +\frac{R_{y}}{R_{z}}\frac{\Delta
_{z}H_{z}^{(m)}}{\Delta z}\ ,  \label{JX} \\
J_{y} &=&\left[ \frac{\Delta _{z}H_{x}^{(m)}}{\Delta z}-\frac{\Delta
_{x}H_{z}^{(m)}}{\Delta x}\right] -\frac{R_{x}}{R_{z}}\frac{\Delta
_{z}H_{z}^{(m)}}{\Delta z}\ .  \label{JY}
\end{eqnarray}%
From these relations and by applying Eq. (\ref{lostosolar}), the $z$\
component of the current density vector in the line-of-sight reference frame
can be derived, which is%
\begin{equation}
J_{z}=\left( J_{Z}-J_{x}R_{x}-J_{y}R_{y}\right) /R_{z}\ ,
\end{equation}%
which can be reduced into%
\begin{eqnarray}
J_{z} &=&\left[ \frac{\Delta _{x}H_{y}^{(m)}}{\Delta x}-\frac{\Delta
_{y}H_{x}^{(m)}}{\Delta y}\right] \\
&&+\frac{R_{x}}{R_{z}}\frac{\Delta _{z}H_{y}^{(m)}}{\Delta z}-\frac{R_{y}}{%
R_{z}}\frac{\Delta _{z}H_{x}^{(m)}}{\Delta z}\ .  \nonumber
\end{eqnarray}%
The second line of the formula accounts for the inclination of the line
formation plane with respect to the line-of-sight. $J_{X}$\ and $J_{Y}$\ can
then be derived by applying Eq. (\ref{lostosolar}) to $J_{x},J_{y}$ and $%
J_{z}$.

In the following, we have plotted the heliographic reference frame
components $J_{X,Y,Z}$\ of the current density vector. For our plots, we 
derived the current density vector coordinates in the heliographic
reference frame from the magnetic field unique vector (after
disambiguation) rotated into the heliographic reference frame by
applying Eq. (\ref{lostosolar}).

\citet{Crouch-etal-09,Crouch-13,Crouch-15} developed a different calculation 
for also accounting for the departure from disk center, which
implies that the line-of-sight is not perpendicular to the line formation
plane. They also obtained that additional terms have to be introduced into
the usual divergence expression. But their formula given in Eqs. (4-5) of %
\citet{Crouch-etal-09,Crouch-13,Crouch-15} is different from our
Eq. (\ref{divergence}) because the corresponding reference
frames are not the same. The divergence of %
\citet{Crouch-etal-09,Crouch-13,Crouch-15} applies spatial
derivatives with respect to the \emph{heliographic} $X$ and $Y$ and \emph{%
line-of-sight} $z$ coordinates, which do not form a rectangular reference
frame, whereas our divergence applies spatial derivatives with
respect to all the line-of-sight $(x,y,z)$ reference frame
coordinates, including the fact that the observed lines are formed
along inclined planes. Both approaches are
different ways to treat the problem of the inclined line-of-sight.

\subsection{The two line formation depth difference}

\label{subsect-depthdiff}

\begin{figure}
\includegraphics[width=8.8cm]{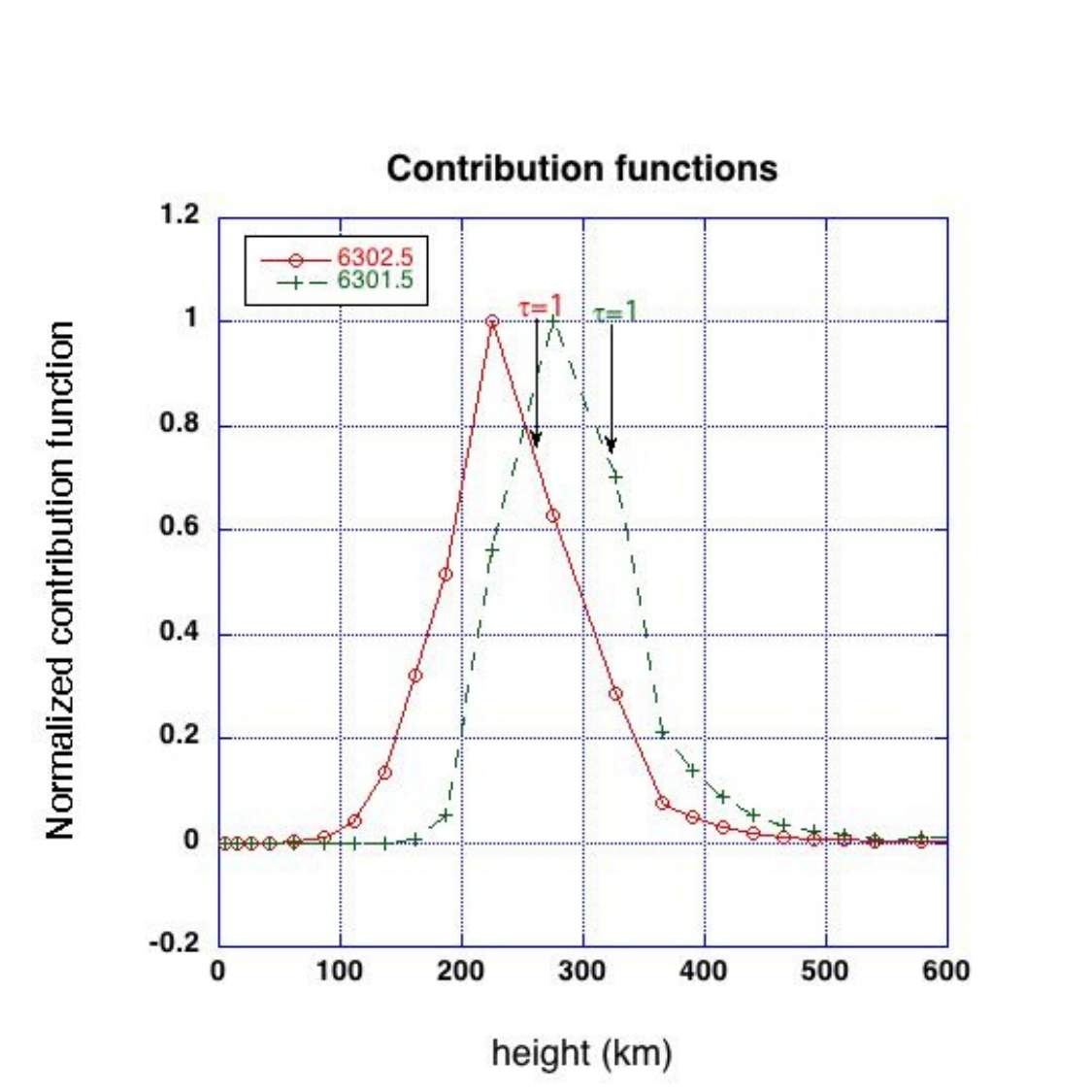}
\caption{Contribution functions of the two lines Fe I 6302.5 \AA\ and 6301.5 \AA\ as a function of the height above $\tau_{5000}=1$. The heights where $\tau=1$, which is the line formation height in the Eddington-Barbier approximation, are indicated by arrows.}
\label{contrib-f}
\end{figure}

The quantity to be determined is the difference in line-of-sight formation
depth between the two lines. As it can be seen in Fig. 4 of %
\citet{Khomenko-Collados-07}, the two lines Fe I 6302.5 \AA\ and 6301.5 \AA\ %
are particularly interesting because their formation depths behave
in a very parallel manner. As also visible in Fig. 4 of %
\citet{Khomenko-Collados-07}, such a parallelism is not the case for the
pair of Fe I 6302.5 and 5250.2 lines. This was the reason
why we discarded 5250.2, though it is more sensitive to the magnetic field
than 6301.5.\ Such a parallelism probably originates in the fact that the
two lines 6302.5 and 6301.5 belong to the same multiplet n. 816 and have
different $gf$\ values, which are respectively $gf=0.180$\ for
6301.5 and $gf=0.0627$\ for 6302.5 from the Kurucz data-basis.
Since differential non-LTE\ effects within multiplets are thought to be very
small, as proven by detailed, multi-level, non-LTE computations,
this implies that the absorption coefficient of 6301.5 is 3 times larger
than the absorption coefficient of 6302.5.\ There is then no doubt that
6301.5 forms higher than 6302.5. In other words, the optical depth
of 6301.5 is almost 3 where the optical depth of 6302.5 is unity. 

The two iron line formation difference in the quiet sun atmosphere was
recently determined by a phase-shift analysis of HINODE observations by %
\citet{Faurobert-etal-09}, who obtained the observed value $63.2\pm
0.9$\ km, which is corroborated by the value of 69 km derived by
the same phase-shift technique applied to theoretical profiles computed with
the non-LTE Uitenbroek's code \citep{Grec-etal-10}.

However, it has to be remarked that the formation height varies
along the line profile. The highest value, which is typically
hundreds of km, is reached at line center, whereas in the
far wings, the formation height is nearly 0 km in the visible range. 
However, in the inversion algorithm, the mean squared difference
between observed and theoretical profiles is computed over the whole
profile. As the Zeeman effect is maximum around line center, this part of
the profile is the most contributing to the field vector determination, so
that it it is quite natural to assign the line center formation height to
this field value. For instance, it is found that the value of $%
1/2\arctan U/Q$\ \ computed at line center provides the field azimuth within
a very good first approximation.

In addition, it has to be remarked that the formation depth has a
certain thickness, which is well represented by the behavior of the
contribution function. We have plotted the contribution functions in Fig. %
\ref{contrib-f} at line center for the two iron lines. This figure
has been obtained by applying a non-LTE polarized radiation
transfer code \citep{Landi-etal-90,Bommier-etal-91} to the Fe I 6302.5 and
6301.5 line formation in the quiet sun reference model atmosphere of %
\citet{Maltby-etal-86}\ and in the absence of a magnetic field.\ It
can be seen in the figure that the formation depth of each line has an
accuracy of the order of $\pm $75 km, which is the halfwidth of the
contribution function. However, the difference $\Delta z=$\ 66 km
between the two lines remains clearly visible all along the profile of the
contribution function. In Fig. \ref{contrib-f} we have also indicated with
arrows the height of $\tau =1$, which we consider as the line
formation depth following the Eddington-Barbier approximation. It is well
visible that this height correctly represents the mean line formation depth.
Fig. 4 of \citet{Khomenko-Collados-07} also shows that the
formation depth difference remains on the order of $\Delta z=$\ 70
km in the quiet sun, whatever the formation depth is in itself.

Finally, in Fig. 4 of \citet{Khomenko-Collados-07} we observed that the
depth difference is larger in active regions than in quiet ones. We thus
applied instead the depth difference of 98 km at disk center for active
regions. This is also the value determined with the non-LTE Uitenbroek's
code \citep{Grec-etal-10} in a previous approximation.

We finally derived the line formation depths from a 
model that we describe below. The temperature, electron pressure, 
and gas pressure were first taken from an atmospheric model. 
We used the \citet{Maltby-etal-86} quiet sun photospheric reference 
model extrapolated downwards beyond $-70$\ km to $-450$\ km below
the $\tau _{5000}=1$\ level (courtesy of IAC). The continuum absorption
coefficient was evaluated as in the MALIP code of \citet{Landi-76}, %
i.e., by including H$^{-}$\ bound-free, H$^{-}$\ free-free, neutral
hydrogen atom opacity, Rayleigh scattering on H atoms, and Thompson
scattering on free electrons. The line absorption coefficient was derived
from the Boltzmann and Saha equilibrium laws taking the two first
ions of element into account. The atomic data were taken from Wiese or Moore
and the partition functions from Wittmann. The iron abundance was assumed to
be 7.60 and the sodium abundance 6.25 in the usual
logarithmic scale where the abundance of hydrogen is 12. A
depth-independent microturbulent velocity field of 1 km/s was introduced.
Finally, departures from LTE in the ionization equilibrium were simulated
for layers above $\tau _{5000}=0.1$ by applying Saha's law with a constant 
radiation temperature of 5100 K instead of the electron temperature
provided by the atmospheric model.

At final step, the line center optical depth grid was scaled to the
continuum optical depth grid by applying the respective
absorption coefficients. We used the continuum optical depth grid 
provided in the atmospheric model and the transfer equation was not 
explicitly solved again. The height of formation of the line center
was then determined as follows. Once obtained the grid of line
center optical depths, the height of formation of the line center %
was located where the optical depth along the line of sight is unity
(Eddington-Barbier approximation), i.e., where $\tau /\mu =1$,
where $\tau $\ is the line center optical depth along the vertical %
and $\mu $\ the cosine of the heliocentric angle $\theta $\ (here supposed
to be 0). As it can be seen in Fig. 5 of \citet{Bruls-etal-91} and in Fig. %
\ref{contrib-f} of the present paper, this conveniently represents the depth
where the contribution function has its maximum.

We thus obtained quiet sun line center formation heights of 262 km for
Fe I 6302.5 and 328 km for Fe I 6301.5 above the $\tau _{5000}=1$\ level, 
leading to a difference of 66 km in excellent agreement 
with the measurements described above, which validates
our computation method.

\subsection{The aspect ratio of the strongly stratified atmosphere}

\label{subsect--aspect}

We were indeed confronted with the problem that the vertical
gradient of the magnetic field $\mu_0\partial H_{z}/\partial z$ was found on the
order of 3 G/km in spot umbrae, whereas the horizontal gradient $\mu_0\partial
H_{x}/\partial x+\mu_0\partial H_{y}/\partial y$ was only on the order of 0.3
G/km.\ This leads to a non-vanishing value of $\func{div}%
\vec{H}$, whatever the signs would be. We then investigated the literature
and found 15 references, which fully confirm these values %
provided by different instruments (groundbased as well as spaceborn),
different inversion methods (SIR or others), and different
spectral lines. The detailed description of these references can be found in %
\citet{Bommier-13}. An observation review was also presented by 
\citet{Balthasar-18}. All concluded to the above cited values. In other
words, a loss of magnetic flux is observed with increasing height, which is
not compensated for by an increase of the horizontal flux. In %
\citet{Bommier-13} and \citet{Bommier-14}, we showed that the lack of
spatial resolution in both transverse and along the line-of-sight
directions cannot be held responsible for the seemingly
non-vanishing observed $\func{div}\vec{H}$. This was based on
mathematical study of the convolution procedure. It was shown that
the divergence computed with averaged quantities is equal to the average of
the local divergences. Accordingly, if the local divergence is zero, the
divergence computed with averaged quantities should be also zero, within the
noise level. The question arised to know if the observed value of $\func{div%
}\vec{H}$\ is an effect of different spatial resolution along the 
different space directions. The effect of the spatial resolution,
horizontal as well as vertical, is a filtering. The mathematical
demonstration that the filtered divergence is the divergence of the filtered
quantity is given in \citet{Bommier-13} and \citet{Bommier-14}. 
\citet{Bommier-14} includes also an easier demonstration in the
spatial Fourier space. The same demonstration applies to the spatial
averaging and the effect of the magnetic filling factor $\alpha $. As
explained in the previous section, we only know the \textquotedblleft 
local average magnetic field \textquotedblright\ $\alpha H$ %
from the measurements, but the average divergence is the divergence of the
averaged field. In addition, \citet{Bommier-14} provides
results of numerical tests devoted to investigate an eventual effect of the
limited spatial resolution (eventually anisotropic). The numerical
tests are all negative. They conclude to a zero divergence computed
by finite differences, when the local divergence is zero. To our opinion,
the negation of this logical proposition is that the non-zero observed value
for the divergence indicates a non-zero local value.\ In \citet{Bommier-15},
we present a discussion about the necessity for $\func{div}\vec{H}$%
\ to be zero. We argue that the existence of magnetic monopoles is not the
only possibility for a non-zero $\func{div}\vec{H}$. After a
discussion about the measurement noise level effect, we present below
another possibility we investigated.

In the TH\'{E}MIS measurements described in \citet{Bommier-etal-07},
which is the validation paper of the UNNOFIT\ inversion method, the
polarimetric noise is assumed to be $1.5\times 10^{-3}$. As
reported in Sect. \ref{subsect--inversion}, this results in an inaccuracy of
10 G for the longitudinal field and 15 G for the transverse field.\ The
field difference observed between and from the two line 6301.5 and 6302.5 is
about 300 G, when the difference in line formation height is about 100 km,
close to disk center. This results in an inaccuracy of 0.2 G/km for the
vertical field gradient observed at disk center. As for the transverse
field, the typical TH\'{E}MIS pixel size is about 500 km \citep{Bommier-13},
which results in an inaccuracy of less than 0.2 G/km for the horizontal
field gradient observed at disk center.\ The total inaccuracy on the
observed $\func{div}\vec{H}$ value results in 0.4 G/km, to be
compared to the non-vanishing value of 2.7 G/km, which results
from the measurements reported in the literature.\ The 
non-vanishing value observed for $\func{div}\vec{H}$ is then
markedly higher than the noise level.

We then investigated how the magnetic field is influenced by the plasma
anisotropy due to the strong stratification due to the gravity and
the density at the star surface, which is responsible for an
\textquotedblleft aspect ratio\textquotedblright\ between horizontal and
vertical typical lengths, respectively denoted as $ l_h $ and $ l_v $. 
In the case of the solar photosphere, by applying
strongly stratified fluid mechanics laws following \citet{Brethouwer-etal-07}%
, we evaluated this aspect ratio to be on the order of $ l_h / l_v = 20 $ %
\citep{Bommier-13,Bommier-14} in the quiet sun photosphere. 
In \citet{Bommier-20}, we showed that what is measured by Zeeman effect
is the magnetic field $\vec{H}$ and not the magnetic induction $\vec{B}$. 
The magnetic induction $\vec{B}$ is divergence-free, which implies that 
$\func{div}\vec{H}=-\func{div}\vec{M}$, where $\vec{M}$ is the
magnetization. The magnetization is linked to the matter.
Accordingly, it scales following the typical lengths introduced above.
It results that once the inverse of
the aspect ratio is applied to scale the different magnetic field
components, the scaled $\func{div}\vec{H}$ vanishes, which enables the
ambiguity resolution as we obtain in this paper. 
An example of this result in the case of NOAA 10808
observed with TH\'{E}MIS on 13 September 2005, is visible in Fig. 3 of %
\citet{Bommier-13}. Although we evaluated the theoretical value of
the aspect ratio on the order of 20 in the quiet sun photosphere, this
ratio may eventually be different in sunpots or plages or active regions%
, which is the case of the regions we treated. This is not so
well-known and sometimes we adjusted the ratio (kept constant in the whole
map) until the ambiguity resolution is consistent in the whole map.
With the word \textquotedblleft consistent\textquotedblright\ here, we mean
that the resulting map does not show any artificial strong current line, 
which would delineate sharp azimuth change due to
ambiguous solution selection change. The resulting map displays at most
homogeneous field directions. We obtain that this inverse scaling is
a necessity for a correct ambiguity resolution. The correctness is obtained
by comparison with the intuitive ambiguity solution, which can be found in
sunpot penumbrae as discussed in the following subsection. 

We applied the Maxwell law 
$\func{curl}\vec{H}=\vec{J}$ for the
current density derivation from the magnetic field vector in different
regions reported in the following.

\subsection{Verification of the ambiguity resolution results}

\label{subsect--verification}

We thus treated 60 maps observed in Fe I 6301/6302 and 26 maps observed in
Na I D with TH\'{E}MIS (in 2010-2013) and 23 HINODE/SOT/SP observations of
active regions\footnote{%
http://lesia.obspm.fr/perso/veronique-bommier/}. For the sunspots the
correctness of the solution can be verified because the spot umbra
polarity is known from circular polarization. This is true even near the
limb, where the solar vertical field at spot center is
nearly transverse with respect to the l.o.s.. The
ambiguity solution is good if it is in agreement with the polarity. It has
to be remarked that for a sunspot observed anywhere on the disk there is 
always a place in it where the ambiguity solution is 
can be derived from the spot polarity. This place is
either in the penumbra, when the spot is observed near the disk center, or
in the umbra, when the spot is observed near the limb. We consider
such cases as\ possibilities of observational proofs\ of our
disambiguation method, and we obtained successful proofs even in a
spot at $\mu =0.42$\ which is $\theta =65%
{{}^\circ}%
$\ from the disk center.

\subsection{Test of the method on theoretical data}

\begin{figure}
\includegraphics[width=5.6cm]{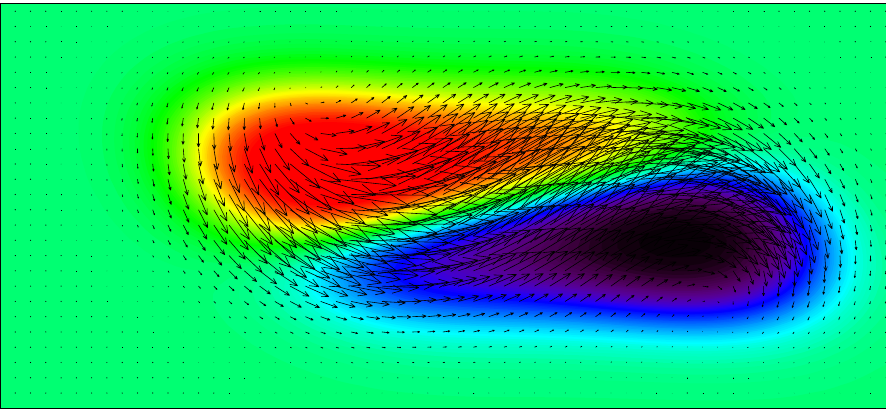}
\includegraphics[width=1.5cm]{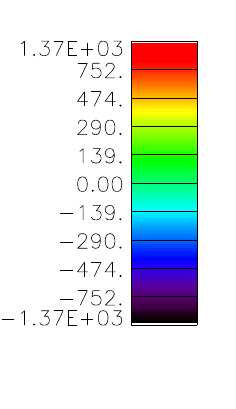}
\includegraphics[width=1.5cm]{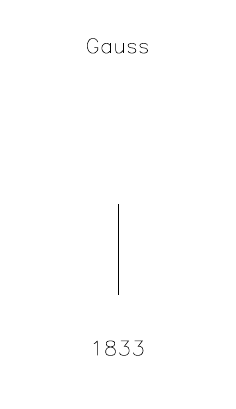}
\includegraphics[width=5.6cm]{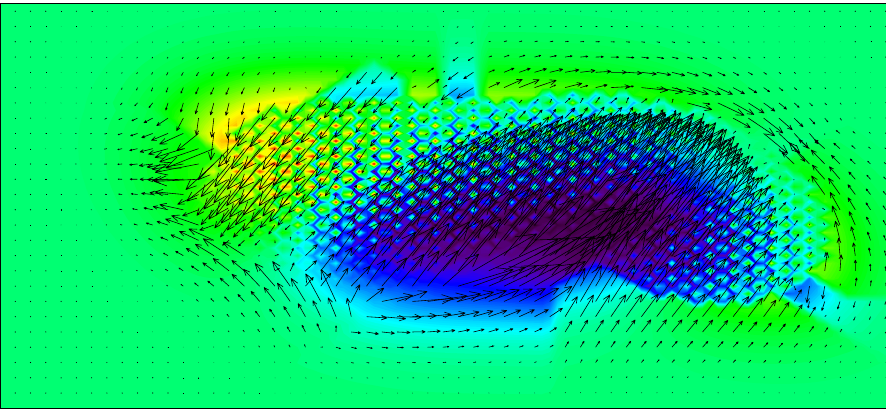}
\includegraphics[width=1.5cm]{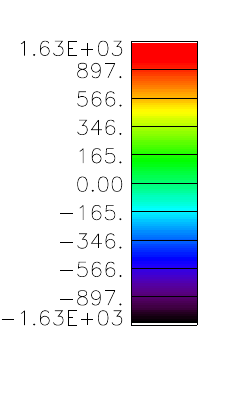}
\includegraphics[width=1.5cm]{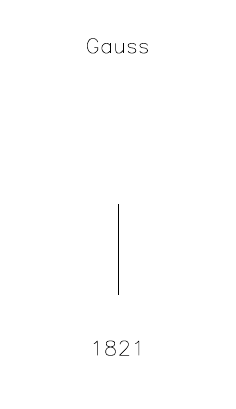}
\includegraphics[width=5.6cm]{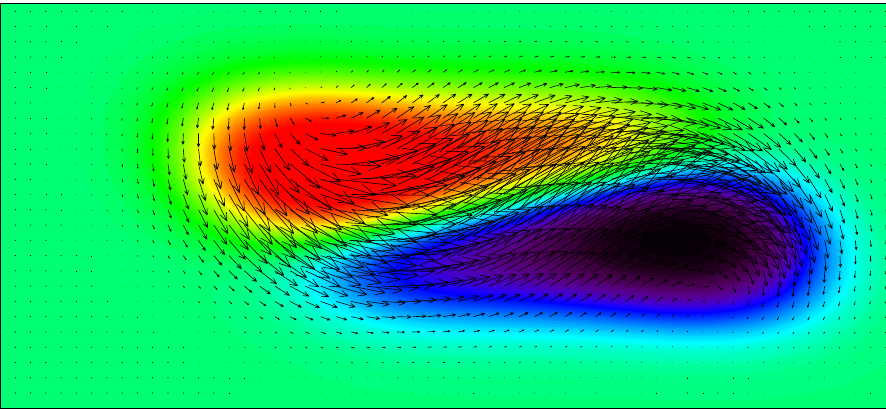}
\includegraphics[width=1.5cm]{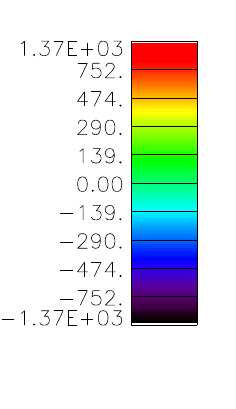}
\includegraphics[width=1.5cm]{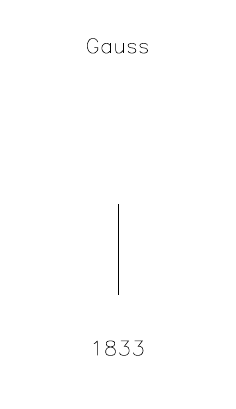}
\caption{Result of test of the ambiguity resolution by our method applied to theoretical data. Top: map of the input magnetic field vector. Middle: ambiguity resolution by applying the usual divergence formula (first line of Eq. (\ref{divergence}) only) to these data, where the region is assumed to be out of disk center and where two lines formed at two different altitudes are supposed to be observed. Bottom: ambiguity resolution by applying Eq. (\ref{divergence}).}
\label{figtest}
\end{figure}

Our method for taking the departure from disk center into account,
which leads to additional terms in the divergence formula Eq. (\ref%
{divergence}), has been tested on theoretical data. Divergence-free
data were prepared by simulating two close sunspots forming
a $\delta $-spot and by applying the magnetic field model described by Eqs.
(8-10) of \citet{Fan-Gibson-04}. Simulation results based on this model have
been used to study $\delta $-spots (as in Fig. 3 of \citet{Leka-etal-05})
and also to test various ambiguity resolution methods, as in %
\citet{Metcalf-etal-06}. The ambiguity of the transverse magnetic field 
in the line-of-sight reference frame was simulated and
submitted to the ME0 ambiguity resolution code of \citet{Leka-etal-09}, 
which was modified by us for taking the vertical magnetic field
gradient $\partial H_{z}/\partial z$\ from 2-line observations
instead from extrapolation as described at the beginning of Sect. %
\ref{subsect--ambiguity}. The 2-line observation had been simulated with
theoretical results at two different depths separated by 70km along the
line-of-sight. The pixel sizes are assumed to be $\Delta x=1160$ km and $%
\Delta y=581$\ km as in our TH\'{E}MIS observations described in %
\citet{Bommier-13}. The $\delta $-spot was assumed to be far from disk
center and located at 30$%
{{}^\circ}%
$\ from the central meridian in longitude and at 45$%
{{}^\circ}%
$\ in latitude. The solar disk center latitude was assumed to be zero.

\ The test result is represented in Fig. \ref{figtest}, where the
theoretical reference field is plotted in the top map. The result
of the disambiguation when the usual divergence formula is applied, 
which is given by the first line of Eq. (\ref{divergence}) only, is
displayed in the middle map. The minimized quantity is in fact $\left\vert 
\func{div}\vec{H}\right\vert +\lambda \left\vert J_{Z}\right\vert $ 
with $\lambda =1$ as in our method. It can be seen that
the amiguity resolution widely fails over the $\delta $-spot. 
When the full Eq. (\ref{divergence}) is applied instead, the
obtained result is displayed in the bottom map of Fig. \ref%
{figtest}, which is in total agreement with the reference field. It
is thus confirmed by numerical test that the second line of Eq. (\ref%
{divergence}) is necessary to resolve the ambiguity by applying a two-line
analysis to maps observed out of disk center as discussed after Eq.
(\ref{divergence-equator}) and in Fig. \ref{partiald}.

\subsection{Test of the resolution code on theoretical data}

\begin{table}
\begin{tabular}{ccccc}
\hline
pixel & line & $\mathcal{M}_{\mathrm{area}}$
& $\mathcal{M}_{H_{\bot}>100\,\mathrm{G}}$ & $\mathcal{M}_{H_{\bot}>500\,\mathrm{G}}$ \\ 
size &  &  &  &  \\ 
\hline
0.3" & lower & 0.99 & 1.00 & 1.00 \\
0.3" & upper & 0.99 & 1.00 & 1.00 \\
0.9" & lower & 0.99 & 1.00 & 1.00 \\
0.9" & upper & 0.99 & 1.00 & 1.00 \\
\hline
\end{tabular}
\caption{Performance metrics for the limited resolution case for our DIVB resolution algorithm applied to the test data of \citet{Leka-etal-09} and \citet{Crouch-13}. The data were provided and the metrics were computed by K.D. Leka's courtesy.}
\label{table1}
\end{table}

\begin{table}
\begin{tabular}{ccccc}
\hline
noise & $\mathcal{M}_{\mathrm{area}}$
& $\mathcal{M}_{H_{\bot}>100\,\mathrm{G}}$ & $\mathcal{M}_{H_{\bot}>500\,\mathrm{G}}$ \\ 
level &  &  &  \\ 
\hline
no &  1.00 & 1.00 & 1.00 \\
low &  0.85 & 0.95 & 1.00 \\
high &  0.73 & 0.83 & 0.90 \\
\hline
\end{tabular}
\caption{Performance metrics for the noise-added case for our DIVB resolution algorithm applied to the test data provided by K.D. Leka's courtesy. These data were similar to that described in \citet{Crouch-13}, but with a two line formation height difference of 890 km along the l.o.s. instead. The metrics were computed by K.D. Leka's courtesy.}
\label{table2}
\end{table}

We tested our ambiguity resolution code DIVB2 built on the above
described method, on the theoretical test data described in %
\citet{Leka-etal-09}. These data were used for testing various ambiguity
resolution methods as in \citet{Leka-etal-09}. The methods by %
\citet{Crouch-etal-09,Crouch-13,Crouch-15}, where magnetic field data at two
different heights are also used, were also submitted to these test data.\ We
used two series of these data. 

The first series was the \textquotedblleft flower\textquotedblright\
data, where the sunspots are simulated with form of flowers, as visible in
Fig. 4 of \citet{Leka-etal-09}. These data were prepared to test the code
robustness against lack of spatial resolution.\ The preparation of the
theoretical data is described in p. 93 of \citet{Leka-etal-09} and also in
p. 111 of \citet{Crouch-13}. We used the theoretical data with averaged
pixel sizes of 0.3" and 0.9". These theoretical data are provided at two
heights separated by 0.18". These data simulate location at disk center,
i.e., the line-of-sight is assumed to be perpendicular to the line formation
planes.\ For these data, we obtained the result metrics listed in Table \ref%
{table1}. The metrics is given by the fraction of pixels with correct
ambiguity resolution in the whole map $\mathcal{M}_{\mathrm{area}}$ %
and the fraction of pixels with correct ambiguity resolution within the
conditions of transverse field $H_{\perp }$ stronger than 100 G and
500 G respectively. As visible in Table \ref{table1}, our code obtains
excellent results for the two resolution cases. 

The second series of data was the simulation represented in Fig. 1
of \citet{Leka-etal-09}. These data were artificially noised by adding them
a theoretical photon noise, at the level of polarimetric accuracy of $%
10^{-3}$ for the low noise case and $10^{-2}$ for the high
noise case, as described in p. 89 of \citet{Leka-etal-09}. The assumed pixel
size was 0.5". The case without any added noise was also treated. These data
are available at two heights. We used data at two heights distant of 890 km
along the l.o.s., whereas \citet{Crouch-13} used data at two heights distant
of the pixel size as described in p. 111 of \citet{Crouch-13}, which is 376
km only. For these data, we obtained the result metrics listed in Table \ref%
{table2}. In this case also, our code obtains very good results. The results
are perfect in the case without any noise. The highest noise level, which is 
$10^{-2}$ in polarization, is rather high and higher than the
current observation polarization inaccuracies. In the case of this high
level noise, we obtain nevertheless rather good results better than 80\%
correct for the high noise level and better than 95\% correct for the low
noise level. As visible in Table \ref{table2}, our worst results locate in
weak magnetic field regions.

This second series of data were assumed to be located out of disk
center at latitude 9$%
{{}^\circ}%
$ South and longitude 36$%
{{}^\circ}%
$ East with respect to the central meridian. This leads to an
heliocentric angle cosine $\mu =\cos \theta =0.80$, which is rather
far from disk center. It has to be remarked that we obtain perfect ambiguity
resolution result with our code in the no noise case as visible in the first
line of Table \ref{table2}. In our code, the disambiguation is performed in
the l.o.s. reference frame and we applied Eq. (\ref{divergence}) to account
for departure from disk center in the case of lines formed along horizontal
planes, which is done in the second line of Eq. (\ref{divergence}). The
success of our resolution of theoretical data again validates our Eq. (\ref%
{divergence}) and, in particular, its second line in the case of lines
formed along horizontal planes and observed out of disk center.

\section{Results: Examples of Vector Maps}

\label{sect--vector-maps}

\begin{figure*}
\includegraphics[width=6cm]{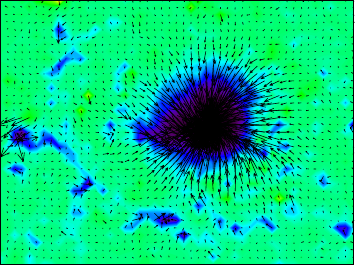}
\includegraphics[width=6cm]{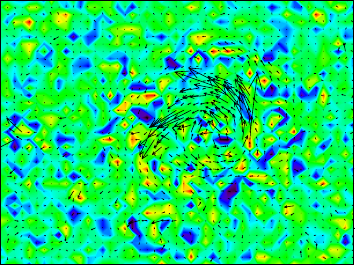}
\includegraphics[width=6cm]{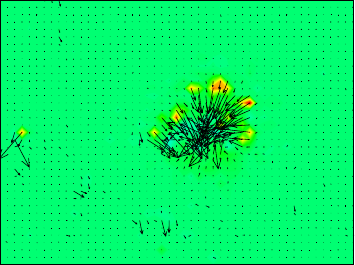}
\caption{Active Region NOAA 11420 observed by HINODE/SOT/SP on 18 February 2012 between 11:08 and 12:04 UT. The ambiguity is resolved by our method. Left: magnetic field vector, vertical component between $-2040$ and $+2040$, horizontal component maximum arrow length 1355 G. Middle: current density vector, vertical component between $-64$ and $+64$, horizontal component typical arrow length 5000 mA/m$^2$. Right: Lorentz force vector, vertical component between $-448$ and $+448$ , horizontal component typical arrow length 800 mN/m$^2$. The spatial resolution was reduced by a factor $5\times 5$ for paper file size purposes. (color figure on-line)}
\label{fighinode1}
\end{figure*}

\begin{figure*}
\includegraphics[width=6cm]{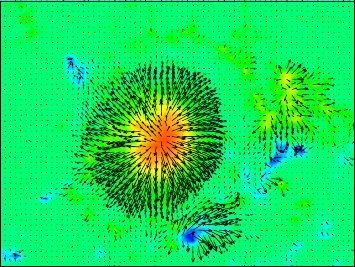}
\includegraphics[width=6cm]{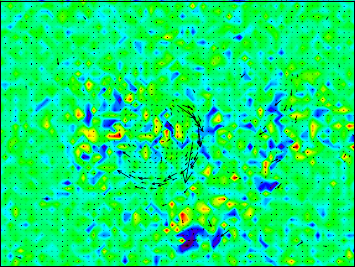}
\includegraphics[width=6cm]{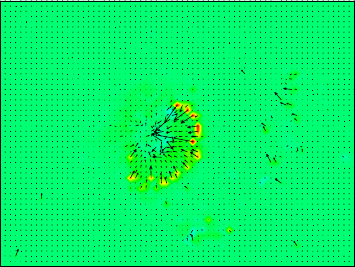}
\caption{Active Region NOAA 11494 observed by HINODE/SOT/SP on 7 June 2012 between 14:15 and 15:05 UT. The ambiguity is resolved by our method. Left: magnetic field vector, vertical component between $-2150$ and $+2150$, horizontal component maximum arrow length 1743 G. Middle: current density vector, vertical component between $-219$ and $+219$, horizontal component typical arrow length 5000 mA/m$^2$. Right: Lorentz force vector, vertical component between $-682$ and $+682$ , horizontal component typical arrow length 800 mN/m$^2$. The spatial resolution was reduced by a factor $5\times 5$ for paper file size purposes. (color figure on-line)}
\label{fighinode2}
\end{figure*}

\begin{figure*}
\includegraphics[width=6cm]{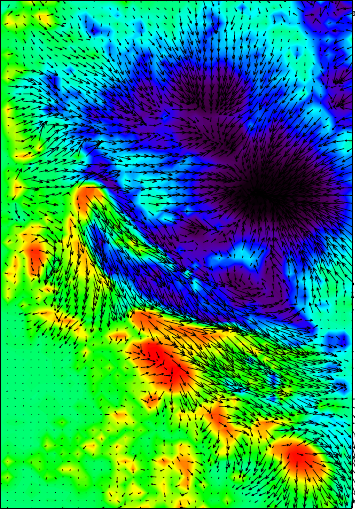}
\includegraphics[width=6cm]{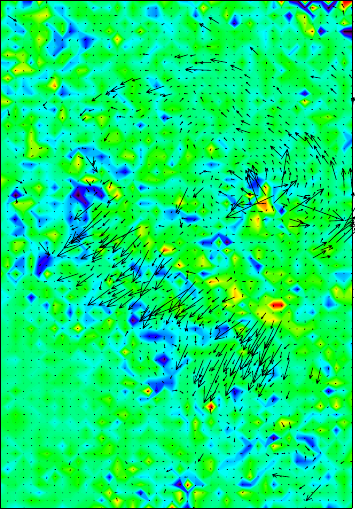}
\includegraphics[width=6cm]{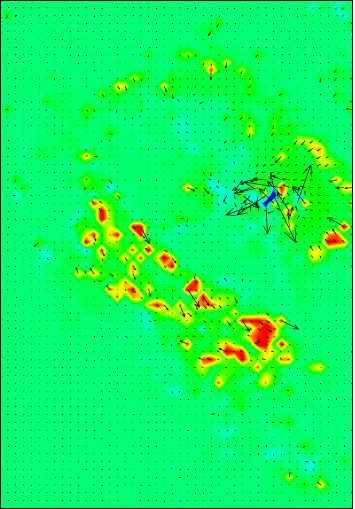}
\caption{Active Region NOAA 11476 observed by HINODE/SOT/SP on 12 May 2012 between 02:38 and 03:11 UT. The ambiguity is resolved by our method. Left: magnetic field vector, vertical component between $-2450$ and $+2450$, horizontal component maximum arrow length 2470 G. Middle: current density vector, vertical component between $-218$ and $+218$, horizontal component typical arrow length 8000 mA/m$^2$. Right: Lorentz force vector, vertical component between $-1040$ and $+1040$ , horizontal component typical arrow length 800 mN/m$^2$. The spatial resolution was reduced by a factor $5\times 5$ for paper file size purposes. (color figure on-line)}
\label{fighinode3}
\end{figure*}

\begin{figure*}
\includegraphics[width=6cm]{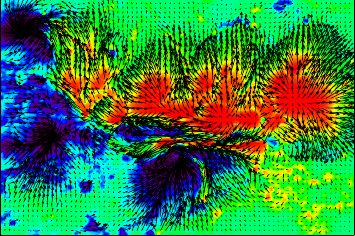}
\includegraphics[width=6cm]{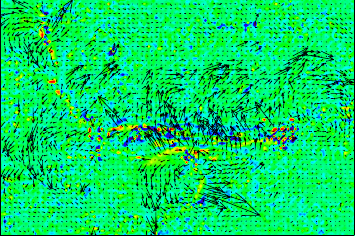}
\includegraphics[width=6cm]{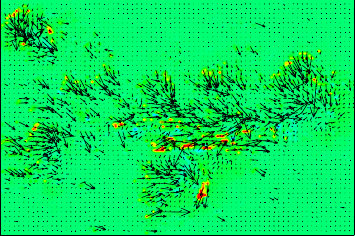}
\caption{Active Region NOAA 11429 observed by HINODE/SOT/SP on 6 March 2012 between 22:10 and 22:43 UT. The ambiguity is resolved by our method. Left: magnetic field vector, vertical component between $-2480$ and $+2480$, horizontal component maximum arrow length 2441 G. Middle: current density vector, vertical component between $-185$ and $+185$, horizontal component typical arrow length 5000 mA/m$^2$. Right: Lorentz force vector, vertical component between $-1180$ and $+1180$ , horizontal component typical arrow length 800 mN/m$^2$. The spatial resolution was reduced by a factor $5\times 5$ for paper file size purposes. (color figure on-line)}
\label{fighinode4}
\end{figure*}

\begin{figure*}
\includegraphics[width=6cm]{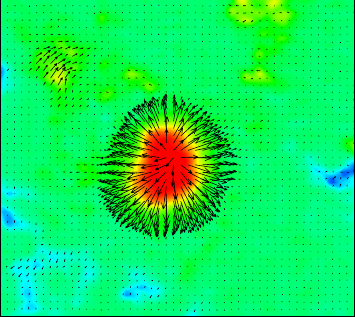}
\includegraphics[width=6cm]{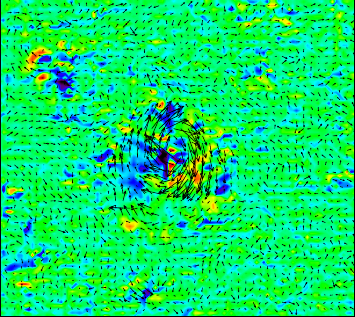}
\includegraphics[width=6cm]{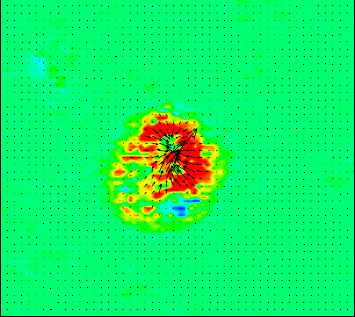}
\caption{Active Region NOAA 11857 observed with THEMIS on 7 October 2013 between 08:57 and 09:22 UT, in the photospheric lines Fe I 6301/6302. The ambiguity is resolved by our method. Left: magnetic field vector, vertical component between $-1790$ and $+1790$, horizontal component maximum arrow length 1070 G. Middle: current density vector, vertical component between $-68$ and $+68$, horizontal component maximum arrow length 746 mA/m$^2$. Right: Lorentz force vector, vertical component between $-56$ and $+56$ , horizontal component maximum arrow length 133 mN/m$^2$. (color figure on-line)}
\label{figthemis1}
\end{figure*}

\begin{figure*}
\includegraphics[width=6cm]{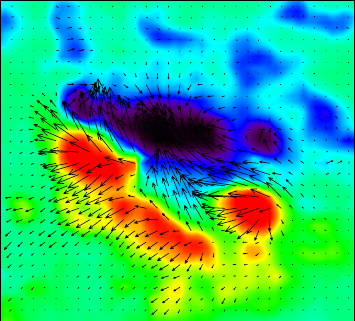}
\includegraphics[width=6cm]{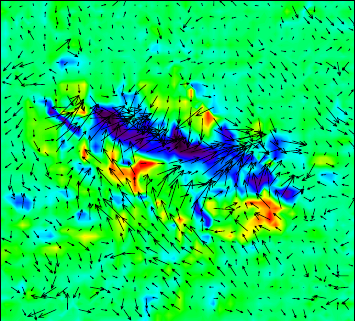}
\includegraphics[width=6cm]{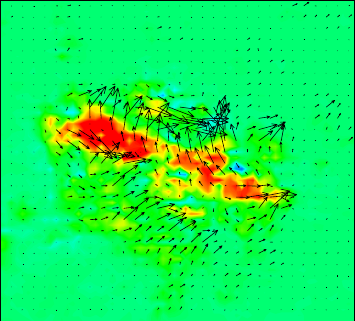}
\caption{Active Region NOAA 11865 observed with THEMIS on 11 October 2013 between 09:39 and 09:59 UT, in the photospheric lines Fe I 6301/6302. The ambiguity is resolved by our method. Left: magnetic field vector, vertical component between $-1450$ and $+1450$, horizontal component maximum arrow length 1320 G. Middle: current density vector, vertical component between $-96$ and $+96$, horizontal component maximum arrow length 655 mA/m$^2$. Right: Lorentz force vector, vertical component between $-83$ and $+83$ , horizontal component maximum arrow length 43 mN/m$^2$. (color figure on-line)}
\label{figthemis3}
\end{figure*}

The results presented below were obtained from 86 maps observed with TH\'{E}%
MIS and 23 maps observed by HINODE/SOT/SP, which are visible in V. Bommier's
personal page\footnote{%
http://lesia.obspm.fr/perso/veronique-bommier}. The magnetic field final
values are also available on-line there. Four examples of HINODE\ data and
two examples of TH\'{E}MIS data are included in Figs. \ref{fighinode1}-\ref%
{figthemis3} below. For accuracy reasons, the HINODE data
pixels were two times binned two by two before the
ambiguity resolution, which results in pixel sizes of $\Delta
x=862 $ km and $\Delta y=928$ km. The TH\'{E}MIS data
pixel size was $\Delta x=794$ km and $\Delta y=725$ km, the binning was not
performed.

The plotted Lorentz force displays the correct direction of the force, but may-be not the 
exact force strength. What is plotted 
in the following figures, 
is $\vec{J} \times \mu_0 \vec{H}$, 
when the Lorentz force is in fact $\vec{J} \times \vec{B}$.
The magnetic induction $\vec{B}$, the magnetic field $\vec{H}$ and the magnetization $\vec{M}$ 
are all parallel, because 
\begin{equation}
\vec{M} = - \frac{(n_e+n_i)k_BT}{B^2} \vec{B}\ ,
\end{equation}
where $n_e$ and $n_i$ are the electron and ion densities respectively, $k_B$ is the Boltzmann constant, and $T$ is the temperature \citep{Delcroix-94}. The minus sign implies that $\vec{M}$ is antiparallel 
to $\vec{B}$, which is plasma diamagnetism. This can be also written as
\begin{equation}
\mu_0 \vec{M} = - \frac{\beta_c}{2} \vec{B}\ ,
\end{equation}
where $\beta_c$ is the plasma $\beta$, ratio of kinetic energy to magnetic energy,
however expressed in terms of charged particles density
\begin{equation}
\beta_c = \frac{(n_e+n_i)k_BT}{B^2/2\mu_0}\ .
\end{equation}
In the usual photospheric electron and ion densities, $\mu_0\vec{M}$ remains negligible with respect to 
$\mu_0\vec{H}$ and $\vec{B}$, which are then equal, and the following figures 
would display the full Lorentz force. However, if the electron density is higher, 
following the suggestion of interior electron thermal escape by \citet{Bommier-20}, 
$\mu_0\vec{M}$ can become non negligible, so that the length of the Lorentz force vectors 
would not be exact in the following figures, even if the vector direction is correct.

\subsection{HINODE data: examples of a single regular spot}

\label{subsect--hinode1}

We systematically observe a strong circular horizontal current component, 
which wraps spots clockwise about a positive polarity spot and
anticlockwise about a negative polarity spot. As a consequence, the
Lorentz force vector is systematically found centripetal with respect to the
spot center. Numerous examples can be found along the web page. We can cite
in particular the active regions NOAA 11420 observed by HINODE/SOT/SP on 18
February 2012 between 11:08 and 12:04 UT (negative polarity spot), %
which is represented in Fig. \ref{fighinode1}, and NOAA 11494\
observed by HINODE/SOT/SP\ on 7 June 2012 between 14:15 and 15:05 UT\
(positive polarity spot), which is represented in Fig. \ref%
{fighinode2}. These examples can be found in the web page with a better
spatial resolution. Similar examples can be found in the web page for
HINODE/SOT/SP observations of single spots on 12 December 2006, 7 September
2011, 1 February 2012, 1 May 2012, and 13 June 2012.

\subsection{HINODE data: examples of a neutral line}

\label{subsect--hinode2}

We often observe a strong horizontal current component, which crosses %
the neutral line.\ Two examples are presented in this paper, which %
can also be found in the web page: NOAA\ 11476 observed by HINODE/SOT/SP on 12
May 2012 between 02:38 and 03:11 UT, which is represented in 
Fig. \ref{fighinode3}, and NOAA\ 11429 observed by HINODE/SOT/SP on
6 March 2012 between 22:10 and 22:43, which is represented in 
Fig. \ref{fighinode4}. One can refer to the web page for a better
spatial resolution. NOAA 11476, which is represented in Fig. \ref%
{fighinode3}, produced numerous C-class flares and NOAA\ 11429,%
\ which is represented in Fig. \ref{fighinode4}, produced an
X-class flare followed by a CME. The example of 7 September 2011, %
which is visible in the web page, is also the case of an active region that
produced an X-class flare. A strong horizontal current component is well
visible across the neutral line.

\subsection{TH\'{E}MIS data: example of a single regular spot}

\label{subsect--THEMIS1}

The example presented in this paper is the case of NOAA\ 11857 observed
with TH\'{E}MIS on 7 October 2013, between 08:57 and 09:22 UT in Fe I
6301/6302. In the case of the photosphere, which is
observed in Fe I 6301/6302 and is represented in %
Fig. \ref{figthemis1}, we find again the current and Lorentz force
characteristics described above, namely circular current about the
spot and centripetal Lorentz force. NOAA 11857 was observed with TH\'{E}%
MIS on 5 October, 7 October (two times,
morning and afternoon), 8 October, and 9 October. All these
examples are visible in the web page. The magnetic field is more rotating
about the spot center at the chromospheric level than at the photospheric 
level, where it is more radial.

\subsection{TH\'{E}MIS data: example of a neutral line}

\label{subsect--THEMIS2}

The example presented in this paper is the case of NOAA\ 11865 observed
with TH\'{E}MIS on 11 October 2013, between 09:39 and 09:59 UT in Fe I
6301/6302. Again, in the photosphere,\ which is
represented in Fig. \ref%
{figthemis3}, we observe a strong horizontal current component across the
neutral line.

\section{Conclusion}

We have shown that observations in two different lines, which belong
to the same multiplet but have different absorption coefficients so
that they are formed at two different depths, like Fe I 6302.5 \AA\ and
6301.5 \AA , enable the resolution of the azimuth ambiguity, which
remain from the Zeeman signal interpretation. The anisotropy of the
strongly stratified plasma of the photosphere has also to be 
accounted for following \citet{Bommier-13} and \citet{Bommier-14}.
As a consequence, two ambiguity-resolved field vector maps are obtained at
two different but close altitudes, which enable the derivation of the
current density full vector via $\func{rot}\vec{H}=\vec{J}$. This
reveals the horizontal component of the current density, which is found
markedly stronger than the better known vertical one, as already
observed by \citet{Puschmann-etal-10b} from HINODE/SOT/SP data.\ We observe
some systematical trends, like circular currents wrapping spots %
clockwise about a positive polarity spot and anticlockwise about a negative
polarity spot and strong horizontal current components, which cross
active region neutral lines. The wrapping direction with respect to the spot
polarity is the same as the average one in Fig. 1 of %
\citet{Puschmann-etal-10b}, who applies completely different methods. 
As a result, the Lorentz force may be computed. 
It is found to be centripetal in sunspots.
The sunspots are thus maintained by the force.

We obtained that the
anisotropic scaling of the real solar data following the aspect ratio in the
strongly stratified medium of the solar photosphere, suggested by %
\citet{Bommier-13} and \citet{Bommier-14}, has to be applied for a
correct resolution of the ambiguity. The correctness can be established by
comparison with the intuitive ambiguity solution, which can be derived in
spot penumbrae from the spot polarity. The necessity of applying the aspect
ratio for solving the ambiguity is another main result of the present work.

This can be done by applying an anisotropic 
scaling factor to the data before disambiguation. For instance, 
by artificially dividing by 10 the pixel sizes before
submission to the AMBIG2 code proposed by \citet{Crouch-13} 
for two lines. This code also requires 
that the observed region be located
close to disk center. Our disambiguation code is able to also disambiguate
data of regions located out of disk center.

We can finally remark that the Na I D1 and D2 lines form another 
favorable line pair, because they are also two lines of the same multiplet 
but with different absorption coefficients.
We evaluated their quiet sun line center formation heights at 533 km for Na I
D1 and 604 km for Na I D2 above the $\tau _{5000}=1$\ level. These lines are
then located in the low chromosphere close to the temperature minimum and
with a difference of 71 km. 
Observing this line pair would open access to 
the low chromosphere where they are formed. 
However, a better spatial resolution for both observations and analysis 
would probably be necessary in such a medium.

In the present analysis, the spatial resolution of both observations 
and analysis is very limited. The Stokes inversion was performed with 
average over the whole observed pixel. The presence of unresolved 
magnetic fields was however roughly accounted for by means of a magnetic 
filling factor $\alpha$, which was introduced and determined in the inversion. 
However, active regions like those studied, which include sunspot umbr\ae\ and 
penumbr\ae\ have structure like dots smaller than the present pixel sizes. 
This concerns the horizontal spatial resolution, but the vertical spatial resolution 
of our study is limited also, because it is limited to the difference in line formation 
depth of two lines, which is found not so far from the vertical height scale. 
As a rough first approach,
we assume a constant formation depth difference for the two lines under interest, 
although our images include various different structures, where this difference
may vary. A better vertical spatial resolution would be attained by applying 
for instance the SIR inversion code \citep{RuizC-delToro-92}.

A series of other effects able to perturb the sunspot magnetic field
measurements is detailed in the above Sect. \ref{subsect--warning}. They are:
velocity or magnetic field-temperature correlations, blending by molecular lines,
straylight from penumbra, departure from a purely hydrostatic equilibrium,
non-horizontal and corrugated $\tau = 1$ surface, non-LTE effects in spectral
lines. Consequently, as long as these and other effects have not been properly tested 
in realistic model atmospheres and with state-of-the-art radiative transfer computations, 
the scaling factor that we introduce must be considered to be a purely empirical factor.

\begin{acknowledgements}
The author is grateful to an anonymous referee for refined suggestions, reported
in this paper.
Some reported observations were made with the
French-Italian telescope TH\'{E}MIS operated by the CNRS on the island of
Tenerife in the Spanish Observatorio del Teide of the Instituto de Astrof%
\'{\i}sica de Canarias. We are grateful to B. Lites for the level-1
HINODE/SOT/SP data. For the data inversion, this work was granted access to
the HPC resources of MesoPSL financed by the Region Ile de France and the
project Equip@Meso (reference ANR-10-EQPX-29-01) of the \textit{%
Investissements d'Avenir} program supervised by the \textit{Agence Nationale
pour la Recherche}.
\end{acknowledgements}

\bibliographystyle{aa}
\bibliography{bommierrefs}

\appendix

\section{Expressions of the divergence and curl when the quantities are
measured along an inclined line formation plane}

\label{Appendix}

We describe below how divergence and curl
have to be computed, when the measurements are made in pixels, 
and when the spectral line used for the measurements is formed at a
certain depth in the stellar atmosphere. This depth defines a formation
plane for the line. In the following we neglect an eventual
thickness of this plane, which we consider as infinitely sharp. We
consider the general case where this plane is inclined with respect to the
line-of-sight for an observation performed out of disk center. The
case of disk center observation is however possible as a particular case.

We denote the heliographic reference frame as $OXYZ$\ and the l.o.s.
reference frame as $Oxyz$ with $Oy$\ solar north oriented.
The point $O$\ where divergence and curl are computed is taken at
longitude-latitude $(L,b)$. The solar radius unit vector $\vec{R}$\ in $O$\
is perpendicular to the the line formation plane, which is tangent
to the solar surface. The $\vec{R}$\ coordinates in the l.o.s.
reference frame are given in Eq. (\ref{R-vector}). In the l.o.s. reference
frame, the line formation plane, which is perpendicular to $\vec{R}$%
\ of coordinates $(R_{x},R_{y},R_{z})$,\ has then for %
general equation%
\begin{equation}
R_{x}x+R_{y}y+R_{z}z=0\ .  \label{plane-eq}
\end{equation}

Let's denote as $\vec{H}^{(1)}$\ the magnetic field vector measured as a
function of $x$\ and $y$\ along the formation plane of the line number (1). $%
\vec{H}^{(1)}$\ depends on the following variables%
\begin{equation}
\vec{H}^{(1)}(x,y,z;R_{x}x+R_{y}y+R_{z}z=0)\ ,
\end{equation}%
where the semicolon means \textquotedblleft such
as\textquotedblright . A similar law holds for the magnetic field vector $%
\vec{H}^{(2)}$\ measured with the second line number (2). In the l.o.s.
reference frame $Oxyz$, we denote as $\Delta x$\ and $\Delta y$\ the
distance between two neighboring pixels along the $Ox$\ and $Oy$\ axes
respectively in the \textquotedblleft sky plane\textquotedblright . %
Let's denote as $\Delta _{x}H_{x}^{(1)}$\ the variation of $H_{x}^{(1)}$\
between two neighboring pixels in $x$\ direction $(i+1,j)$\ and $(i,j)$\
distant of $\Delta x$\ in the image\ 
\begin{equation}
\Delta _{x}H_{x}^{(1)}=H_{x}^{(1)}(i+1,j)-H_{x}^{(1)}(i,j)\ ,
\end{equation}%
and analogously 
\begin{equation}
\Delta _{y}H_{y}^{(1)}=H_{y}^{(1)}(i,j+1)-H_{y}^{(1)}(i,j)\ .
\end{equation}%
For the second line one can similarly define $\Delta
_{x}H_{x}^{(2)} $\ and $\Delta _{y}H_{y}^{(2)}$. The magnetic field may be
averaged between the two lines, with the acute angle method when the
ambiguity is not resolved, and in this case we denote the average as $\Delta
_{x}H_{x}^{(m)}$\ and $\Delta _{y}H_{y}^{(m)}$. As for the $z$\ variation,
it involves the two lines as%
\begin{equation}
\Delta _{z}H_{z}^{(m)}=H_{z}^{(1)}(i,j)-H_{z}^{(2)}(i,j)\ .
\end{equation}

\subsection{Calculation of the divergence}

The mathematical expression of $\func{div}\vec{H}$\ is frame-independent 
\begin{equation}
\func{div}\vec{H}=\frac{\partial H_{x}}{\partial x}+\frac{\partial H_{y}}{%
\partial y}+\frac{\partial H_{z}}{\partial z}\ .
\end{equation}%
However, a problem arises when it is discretized in the above
described system of line formation planes. $\Delta
_{x}H_{x}^{(1)}/\Delta x$,\ which could a priori be considered as
the $\partial H_{x}/\partial x$\ contribution, includes also a variation of $%
H_{x}^{(1)}$\ along the $Oz$\ axis because the magnetic field is
measured along the line formation plane. As a consequence, $\Delta
_{x}H_{x}^{(1)}/\Delta x$\ cannot finally be considered as %
an approximation of $\partial H_{x}/\partial x$ because $\partial
H_{x}/\partial x$\ is the variation of $H_{x}$\ with $x$\ at constant $y$\
and $z$ following the mathematical definition.

The present Appendix is devoted to account for this difficulty by applying
the Ostrogradski's theorem, which is that the volume integrated divergence
equals the flux of the quantity through the surface that borders the volume.
We apply this theorem to the small volume delineated by four neighboring
pixels separated by $\Delta x$\ and $\Delta y$\ in the sky plane $xOy$\ but
located in the line formation plane, and the corresponding four other pixels
for the second line formed in another formation plane separated from the
first one by the distance $\Delta z$\ along the line-of-sight. In the
following we detail three different derivations from the simplest to the
most complex. We consider first an observation located at the solar equator
without inclination of the solar rotation axis\ (Sect. \ref{app -- subsect
-- equator}). We then generalize this demonstration to the general case
(Sect. \ref{app -- subsect -- general I}) and we add a more elegant
derivation leading to the same result as a third part (Sect. \ref{app --
subsect -- general II}).

\subsubsection{Case of a region located at solar equator}

\label{app -- subsect -- equator}

\begin{figure}
\begin{center}
\includegraphics[width=8.8cm]{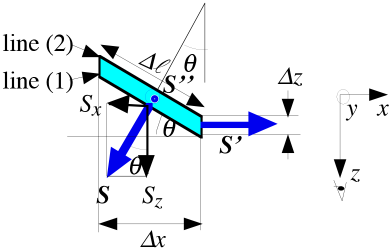}
\end{center}
\caption{Reference frame for the divergence evaluation: case of an observation at the solar equator. The heliocentric $\theta$ angle is oriented by the $Oy$ axis and is negative in the figure. The two formation planes corresponding to the two lines (1) and (2) are indicated by their respective number. The unit $S$ vector is indeed the same as the unit $R$ vector along the local solar radius.}
\label{frame2D}
\end{figure}

We first assume that $O$\ is located on solar equator and that the
Sun rotation axis lies in the plane of the sky, i.e. that $b_{0}=0$. The
heliocentric angle is denoted as $\theta $. A section of the small volume
introduced above is represented in Fig. \ref{frame2D}. For each side of the
volume, a unit vector perpendicular to the surface and pointing towards the
exterior of the volume can be plotted. Three of these vectors are plotted in
Fig. \ref{frame2D}, $\vec{S}$, $\vec{S}^{\prime }$,\ and $\vec{S}%
^{\prime \prime }$. One has to calculate the flux of the magnetic field
through each elementary surface. Consider first the surface element
associated to the $\vec{S}$\ vector. This surface element is part of the
formation plane of line number (1), whereas $\vec{S}^{\prime }$\ and $\vec{S}%
^{\prime \prime }$\ join the two line formation planes. In the l.o.s.
reference frame $Oxyz$, the $\vec{S}$\ unit vector has the following
coordinates%
\begin{equation}
\vec{S}\left\{ 
\begin{array}{l}
S_{x}=\sin \theta \\ 
S_{y}=0 \\ 
S_{z}=\cos \theta%
\end{array}%
\right. \ .
\end{equation}%
As a consequence, the flux of the magnetic field through this
surface per surface unit is%
\begin{equation}
\vec{H}^{(1)}\cdot \vec{S}=H_{z}^{(1)}\cos \theta +H_{x}^{(1)}\sin \theta \ .
\end{equation}%
As visible in Fig. \ref{frame2D}, the elementary surface value 
is $\Delta \ell \Delta y$, where%
\begin{equation}
\Delta \ell =\frac{\Delta x}{\cos \theta }\ .  \label{delta-l}
\end{equation}%
The flux through this elementary surface is then finally%
\begin{equation}
\vec{H}^{(1)}\cdot \vec{S}\ \Delta \ell \Delta
y=(H_{z}^{(1)}+H_{x}^{(1)}\tan \theta )\ \Delta x\Delta y\ .
\end{equation}%
The flux through the elementary surface associated to $\vec{S}^{\prime }$\ is%
\begin{equation}
\vec{H}^{(m)}\cdot \vec{S}^{\prime }\ \Delta z\Delta y=H_{x}^{(m)}\Delta
z\Delta y\ ,
\end{equation}%
where the superscript $(m)$\ denotes the average between the two lines $(1)$%
\ and $(2)$. Finally, the flux through the third elementary surface $\vec{S}%
^{\prime \prime }$\ is%
\begin{equation}
\vec{H}^{(m)}\cdot \vec{S}^{\prime \prime }\ \Delta \ell \Delta z\cos \theta
=H_{y}^{(m)}\Delta x\Delta z\ .
\end{equation}%
As the elementary volume is $\Delta \ell \Delta y\Delta z\cos \theta =\Delta
x\Delta y\Delta z$, one has finally in application of Ostrogradski's theorem%
\begin{eqnarray}
&&\func{div}\vec{H}\ \Delta x\Delta y\Delta z  \nonumber \\
&=&\left[ (H_{z}^{(1)}-H_{z}^{(2)})+(H_{x}^{(1)}-H_{x}^{(2)})\tan \theta %
\right] \ \Delta x\Delta y \\
&&+\left[ H_{x}^{(m)}(x+\Delta x,y)-H_{x}^{(m)}(x,y)\right] \ \Delta z\Delta
y  \nonumber \\
&&+\left[ H_{y}^{(m)}(x,y+\Delta y)-H_{y}^{(m)}(x,y)\right] \ \Delta x\Delta
z\ ,  \nonumber
\end{eqnarray}%
which results in Eq. (\ref{divergence-equator}) of the paper. The term in $%
\tan \theta $\ accounts for the line formation plane inclination with
respect to the line-of-sight.

\subsubsection{Generalization}

\label{app -- subsect -- general I}

\begin{figure}
\begin{center}
\includegraphics[width=8.8cm]{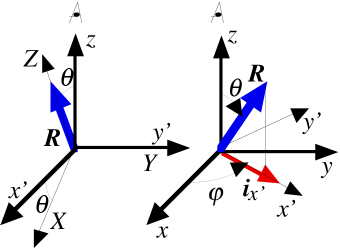}
\end{center}
\caption{Reference frame for the divergence evaluation: general case.}
\label{frame3D}
\end{figure}

One considers now a point $O$\ on the solar surface with longitude $L$\ and
latitude $b$. The solar radius in $O$\ denoted as $\vec{R}$\ is
perpendicular to the line formation plane. The $\vec{R}$\ coordinates in
terms of $(L,b)$\ are given in Eq. (\ref{R-vector}). There is a common
perpendicular to this radius $R$\ and to the l.o.s. $Oz$. We denote as $%
Oy^{\prime }=OY$\ this common perpendicular, which is represented in the
left part of Fig. \ref{frame3D}. In this section, the $OXYZ$\ reference
frame is different from the rest of the paper, because the common
perpendicular $OY$\ is not necessarily the solar meridian. $Oy^{\prime }$\
is also different from the $Oy$\ axis of the l.o.s. reference frame,
therefore we use a different notation. In the $Ox^{\prime }y^{\prime }z$\
reference frame the geometry is the same as in the equatorial case studied
above, so that its final result may be applied%
\begin{eqnarray}
\func{div}\vec{H} &=&\frac{\Delta _{x^{\prime }}H_{x^{\prime }}^{(m)}}{%
\Delta x^{\prime }}+\frac{\Delta _{y^{\prime }}H_{y^{\prime }}^{(m)}}{\Delta
y^{\prime }}+\frac{\Delta _{z}H_{z}^{(m)}}{\Delta z} \\
&&+\tan \theta \frac{\Delta _{z}H_{x^{\prime }}^{(m)}}{\Delta z}\ , 
\nonumber
\end{eqnarray}%
where the $\Delta $'s are taken along each respective axis.

One has now to rotate this result into the l.o.s. reference frame
represented in the right part of Fig. \ref{frame3D}. In this reference
frame, the $\vec{i}_{x^{\prime }}$\ unit vector along $Ox^{\prime }$\ has 
the following coordinates%
\begin{equation}
\vec{i}_{x^{\prime }}\left\{ 
\begin{array}{l}
\cos \varphi \\ 
\sin \varphi \\ 
0%
\end{array}%
\right. \ .
\end{equation}%
Accordingly,%
\begin{equation}
H_{x^{\prime }}^{(m)}=H_{x}^{(m)}\cos \varphi +H_{y}^{(m)}\sin \varphi \ .
\end{equation}%
One has by rotational invariance%
\begin{equation}
\frac{\Delta _{x^{\prime }}H_{x^{\prime }}^{(m)}}{\Delta x^{\prime }}+\frac{%
\Delta _{y^{\prime }}H_{y^{\prime }}^{(m)}}{\Delta y^{\prime }}=\frac{\Delta
_{x}H_{x}^{(m)}}{\Delta x}+\frac{\Delta _{y}H_{y}^{(m)}}{\Delta y}\ ,
\end{equation}%
so that finally%
\begin{eqnarray}
\func{div}\vec{H} &=&\frac{\Delta _{x}H_{x}^{(m)}}{\Delta x}+\frac{\Delta
_{y}H_{y}^{(m)}}{\Delta y}+\frac{\Delta _{z}H_{z}^{(m)}}{\Delta z}+\frac{1}{%
\cos \theta }  \label{div-app-gen-1} \\
&&\times \left[ \frac{\Delta _{z}H_{x}^{(m)}}{\Delta z}\sin \theta \cos
\varphi +\frac{\Delta _{z}H_{y}^{(m)}}{\Delta z}\sin \theta \sin \varphi %
\right] \ .  \nonumber
\end{eqnarray}%
The $\vec{R}$\ vector coordinates can also be expressed as%
\begin{equation}
\vec{R}\left\{ 
\begin{array}{l}
\sin \theta \cos \varphi \\ 
\sin \theta \sin \varphi \\ 
\cos \theta%
\end{array}%
\right. \ .  \label{R-vector-theta-phi}
\end{equation}%
Accordingly, the above Eq. (\ref{div-app-gen-1}) simply
results in Eq. (\ref{divergence}) of the paper. The second line accounts for
the line formation plane inclination with respect to the line-of-sight

\subsubsection{General case}

\label{app -- subsect -- general II}

\begin{figure}
\begin{center}
\includegraphics[width=8.8cm]{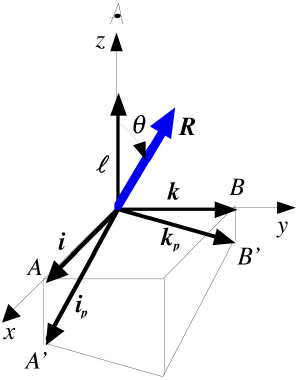}
\end{center}
\caption{Projection of the unit vectors of the line-of-sight reference frame onto the line formation plane, which is perpendicular to the solar radius unit vector $\vec{R}$.}
\label{gen-frame3D}
\end{figure}

Let's introduce the l.o.s. basic unit vectors $\vec{i}$, $\vec{k}$\ and $%
\vec{\ell}$\ (l.o.s. vector), and let's consider their projections $\vec{i}%
_{p}$\ and $\vec{k}_{p}$ along $Oz$\ onto the line formation plane
perpendicular to the solar radius unit vector $\vec{R}$\ (see Fig. \ref%
{gen-frame3D}). $\vec{i}_{p}$\ and $\vec{k}_{p}$\ are not unit vectors and
have to be evaluated as follows. Let's denote by $A$\ the extremity of the $%
\vec{i}$\ vector, and by $A^{\prime }$\ the extremity of the $\vec{i}_{p}$\
vector. $A^{\prime }$\ is the projection of $A$\ along $Oz$. The $AA^{\prime
}$\ line of Fig. \ref{gen-frame3D} is defined by the intersection of the two
planes\ of equations $x=1$\ and $y=0$, respectively. $A^{\prime }$\ is the
intersection of the $AA^{\prime }$\ line with the line formation plane
perpendicular to $R$\ whose equation is given above in Eq. (\ref{plane-eq}).
The coordinates of $\vec{i}_{p}$\ are then%
\begin{equation}
\vec{i}_{p}\left\{ 
\begin{array}{c}
1 \\ 
0 \\ 
-\dfrac{R_{x}}{R_{z}}%
\end{array}%
\right. \ .
\end{equation}%
Analogously for the $\vec{k}$\ and $\vec{k}_{p}$\ vectors, the $BB^{\prime }$%
\ line of Fig. \ref{gen-frame3D} is defined by the intersection of the two
planes\ of equations $x=0$\ and $y=1$ respectively and the
coordinates of $\vec{k}_{p}$\ are obtained by intersection with the line
formation plane given in Eq. (\ref{plane-eq})%
\begin{equation}
\vec{k}_{p}\left\{ 
\begin{array}{c}
0 \\ 
1 \\ 
-\dfrac{R_{y}}{R_{z}}%
\end{array}%
\right. \ .
\end{equation}

We now apply the Ostrogradski's theorem. The elementary volume is obtained
by projecting the elementary surface $\Delta x\Delta y$\ along $Oz$\ onto
the map plane, and by giving it a $\Delta z$\ thickness along $Oz$.
As visible in Fig. \ref{gen-frame3D}, the surface of the projected $\Delta
x\Delta y$\ is $\vec{i}_{p}\ \Delta x\times \vec{k}_{p}\ \Delta y$, which
has for value $\Delta x\Delta y/R_{z}=\Delta x\Delta y/\cos \theta $,\ 
where $\theta $\ is the heliocentric angle. The surfaces of the
other sides of the elementary volume are respectively $\Delta x\Delta z$\
and $\Delta y\Delta z$\ (they are obtained by taking the product of $\vec{i}%
_{p}$\ or $\vec{k}_{p}$\ with $\vec{\ell}$, which is the unit
vector along the l.o.s.). The divergence is then obtained as%
\begin{eqnarray}
\func{div}\vec{H}\ \Delta x\Delta y\Delta z &=&\left[ \vec{H}^{(m)}(z+\Delta
z)-\vec{H}^{(m)}(z)\right]  \nonumber \\
&&\cdot \left[ \vec{i}_{p}\ \Delta x\times \vec{k}_{p}\ \Delta y\right] 
\nonumber \\
&&+\left[ \vec{H}^{(m)}(x+\Delta x)-\vec{H}^{(m)}(x)\right]  \nonumber \\
&&\cdot \left[ \vec{k}_{p}\ \Delta y\times \vec{\ell}\ \Delta z\right] 
\nonumber \\
&&+\left[ \vec{H}^{(m)}(y+\Delta y)-\vec{H}^{(m)}(y)\right]  \nonumber \\
&&\cdot \left[ \vec{\ell}\ \Delta z\times \vec{i}_{p}\ \Delta x\right]
\end{eqnarray}%
leading to the expression given in Eq. (\ref{divergence}). For clarity, the
repeated and unchanged indices have been omitted in $\vec{H}^{(m)}$.

\subsection{Calculation of the curl}

The curl is similarly obtained by applying the Stokes'
theorem, which is that the flux of the curl through an elementary surface
equals the \textquotedblleft circulation\textquotedblright\ or work of the
vector under study along the frontier of the surface, which is
oriented as the surface itself. The elementary surfaces are the different
faces of the elementary volume introduced above for the divergence.
Analogously to the divergence derivation, we present first below a detailed
calculation in the particular case of solar equator (Sect. \ref{app
-- subsect -- equator -- curl}) and then the general case (Sect. %
\ref{app -- subsect -- general II -- curl}).

\subsubsection{Case of a region located at solar equator}

\label{app -- subsect -- equator -- curl}

\begin{figure}
\begin{center}
\includegraphics[width=8.8cm]{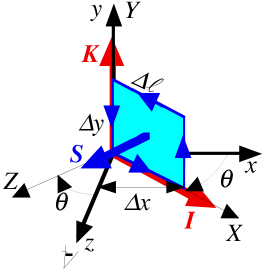}
\end{center}
\caption{Elementary surface for the calculation of $J_Z$ by applying the Stokes theorem to the circulation of the magnetic field vector along the oriented frontier of the surface. This figure is another view of Fig. \ref{frame2D}. Accordingly, the heliocentric $\theta$ angle is oriented by the $Oy$ axis and is negative in the figure. The unit $S$ vector is indeed the same as the unit $R$ vector along the local solar radius.}
\label{curl2D}
\end{figure}

We consider the calculation of $J_{Z}$\ from the circulation of $\vec{H}$\
along the oriented frontier of the elementary surface perpendicular to $OZ$.
We have represented this elementary surface in Fig. \ref{curl2D}, which is
another view of Fig. \ref{frame2D}. The circulation of $\vec{H}$\ about the
oriented surface is given by the scalar products%
\begin{eqnarray}
&&\left[ \vec{H}^{(m)}(y)-\vec{H}^{(m)}(y+\Delta y)\right] \cdot \vec{I}%
\Delta \ell \\
&&+\left[ \vec{H}^{(m)}(x+\Delta x)-\vec{H}^{(m)}(x)\right] \cdot \vec{K}%
\Delta y\ .  \nonumber
\end{eqnarray}%
For clarity, the repeated and unchanged indices have been omitted in $\vec{H}%
^{(m)}$. The unit vector $\vec{I}$\ along $OX$\ has the following
coordinates in the l.o.s. reference frame $Oxyz$%
\begin{equation}
\vec{I}\left\{ 
\begin{array}{l}
\cos \theta \\ 
0 \\ 
-\sin \theta%
\end{array}%
\right. \ ,
\end{equation}%
whereas $\vec{K}$\ is%
\begin{equation}
\vec{K}\left\{ 
\begin{array}{l}
0 \\ 
1 \\ 
0%
\end{array}%
\right. \ .
\end{equation}%
As a consequence, the circulation of $\vec{H}$\ is%
\begin{eqnarray}
&&\left[ H_{x}^{(m)}(y)\cos \theta -H_{z}^{(m)}(y)\sin \theta \right] \Delta
\ell  \nonumber \\
&&+\left[ H_{y}^{(m)}(x+\Delta x)\right] \Delta y \\
&&-\left[ H_{x}^{(m)}(y+\Delta y)\cos \theta -H_{z}^{(m)}(y+\Delta y)\sin
\theta \right] \Delta \ell  \nonumber \\
&&-\left[ H_{y}^{(m)}(x)\right] \Delta y\ .  \nonumber
\end{eqnarray}%
This circulation equals the flux of $\vec{J}$\ through the surface, which is 
$J_{Z}$\ times the surface value, which is $\Delta \ell \Delta
y=\Delta x\Delta y/\cos \theta $\ following Eq. (\ref{delta-l}). %
This can be written as%
\begin{eqnarray}
J_{Z} &=&\cos \theta \left[ \frac{\Delta _{x}H_{y}^{(m)}}{\Delta x}-%
\frac{\Delta _{y}H_{x}^{(m)}}{\Delta y}\right]  \nonumber \\
&&+\sin \theta \frac{\Delta _{y}H_{x}^{(m)}}{\Delta y}\ ,
\end{eqnarray}%
which is Eq. (\ref{rotationnel}) of the paper in the case of the
surface located at equator, where the coordinates of the $\vec{R}$\
vector are given by Eq. (\ref{R-vector-theta-phi}) with $\varphi =0$%
.

\subsubsection{General case}

\label{app -- subsect -- general II -- curl}

Referring now to Fig. \ref{gen-frame3D}, the sides of the elementary volume
are respectively $\vec{i}_{p}\Delta x$, $\vec{k}_{p}\Delta y$\ and $\vec{\ell%
}\ \Delta z$. The circulation of $\vec{H}$\ is given by its
scalar product with each of these elementary vectors, eventually
reversed following the surface orientation. This leads to%
\begin{eqnarray}
J_{Z}\ \frac{\Delta x\Delta y}{R_{z}} &=&\vec{J}\cdot \left[ 
\vec{i}_{p}\ \Delta x\times \vec{k}_{p}\ \Delta y\right] \\
&=&\left[ \vec{H}^{(m)}(y)-\vec{H}^{(m)}(y+\Delta y)\right] \cdot \vec{i}%
_{p}\ \Delta x  \nonumber \\
&&+\left[ \vec{H}^{(m)}(x+\Delta x)-\vec{H}^{(m)}(x)\right] \cdot \vec{k}%
_{p}\ \Delta y  \nonumber
\end{eqnarray}%
which results in the expression given in Eq. (\ref{rotationnel}). The flux
through the two other faces, which are respectively perpendicular to the $Ox$%
\ and $Oy$\ of the line-of-sight reference frame, gives the current
components along those axes%
\begin{eqnarray}
J_{y}\ \Delta x\Delta z\ &=&\vec{J}\cdot \left[ \vec{\ell}\
dz\times \vec{i}_{p}\ \Delta x\right] \\
&=&\left[ \vec{H}^{(m)}(z+\Delta z)-\vec{H}^{(m)}(z)\right] \cdot \vec{i}%
_{p}\ \Delta x  \nonumber \\
&&+\left[ \vec{H}^{(m)}(x)-\vec{H}^{(m)}(x+\Delta x)\right] \cdot \vec{\ell}%
\ \Delta z  \nonumber
\end{eqnarray}%
and%
\begin{eqnarray}
J_{x}\ \Delta y\Delta z\ &=&\vec{J}\cdot \left[ \vec{k}_{p}\
\Delta y\times \vec{\ell}\ \Delta z\right] \\
&=&\left[ \vec{H}^{(m)}(z)-\vec{H}^{(m)}(z+\Delta z)\right] \cdot \vec{k}%
_{p}\ \Delta y  \nonumber \\
&&+\left[ \vec{H}^{(m)}(y+\Delta y)-\vec{H}^{(m)}(y)\right] \cdot \vec{\ell}%
\ \Delta z\ .  \nonumber
\end{eqnarray}%
This results in the expressions given in Eqs. (\ref{JX}-\ref{JY}).
Again, for clarity, the repeated and unchanged indices have been omitted in $%
\vec{H}^{(m)}$.

These three derived current density vector coordinates are not all taken in
the same reference frame. The frame rotation is fully accounted for in Sect. %
\ref{subsect-divrot}, in which all the coordinates of the current density
vector are finally derived in both reference frames.

\end{document}